\documentclass[journal]{IEEEtran}

\IEEEoverridecommandlockouts

\ifCLASSINFOpdf
  \usepackage[pdftex]{graphicx}
 \else
  \usepackage[dvips]{graphicx}
\fi

\usepackage[cmex10]{amsmath}
\usepackage{amssymb,color}

\usepackage{array}

\usepackage{amssymb}
\usepackage{enumitem}
\usepackage{verbatim}
\usepackage{bbold}

\usepackage{bbm}

\newtheorem{thm}{Theorem}
\newtheorem{remark}{Remark}
\newtheorem{cor}{Corollary}
\newtheorem{lem}{Lemma}

\newtheorem{con}{Conjecture}

\newcommand{\ket}[1]{|{#1}\rangle}

\newcommand{\bra}[1]{\langle{#1}|}

\newcommand{\braket}[2]{\langle{#1}|{#2}\rangle}

\newcommand{\E}{\mathcal{E}}
\newcommand{\Ech}{\E_{\sf ch}}

\newcommand{\Hil}{\mathcal{H}}

\newcommand{\Tr}{\text{Tr}}
\newcommand{\beq}{\begin{equation}}
\newcommand{\eeq}{\end{equation}}

\makeatletter
\newcommand{\rmnum}[1]{\romannumeral #1}
\newcommand{\Rmnum}[1]{\expandafter\@slowromancap\romannumeral #1@}
\makeatother

\begin{document}
%
\title{Superadditivity of Quantum Channel Coding Rate with Finite Blocklength Joint Measurements}

\author{Hye Won Chung,~Saikat Guha~and~Lizhong~Zheng
\thanks{Hye Won Chung (hyechung@umich.edu) was with the EECS department at MIT and is currently with the EECS department at the University of Michigan. Lizhong Zheng (lizhong@mit.edu) is with the EECS department at MIT. Saikat Guha (sguha@bbn.com) is with the Quantum Information Processing (QuIP) group at Raytheon BBN Technologies. This paper was presented in part at the 2013 IEEE Allerton conference in Monticello, IL, USA~\cite{chung2013eecs} and the 2014 IEEE International Symposium on Information Theory (ISIT)
in Honolulu, HI, USA~\cite{chung2014super}. This research was partially supported by the DARPA Information in a Photon (InPho) program under contract number HR0011-10-C-0159.


 }}

\maketitle

\begin{abstract}
The maximum rate at which classical information can be reliably transmitted per use of a quantum channel strictly increases in general with $N$, the number of channel outputs that are detected jointly by the quantum joint-detection receiver (JDR). This phenomenon is known as superadditivity of the maximum achievable information rate over a quantum channel. We study this phenomenon for a pure-state classical-quantum (cq) channel and provide a lower bound on $C_N/N$, the maximum information rate when the JDR is restricted to making joint measurements over no more than $N$ quantum channel outputs, while allowing arbitrary classical error correction. We also show the appearance of a superadditivity phenomenon---of mathematical resemblance to the aforesaid problem---in the channel capacity of a classical discrete memoryless channel (DMC) when a concatenated coding scheme is employed, and the inner decoder is forced to make hard decisions on $N$-length inner codewords. Using this correspondence, we develop a unifying framework for the above two notions of superadditivity, and show that for our lower bound to $C_N/N$ to be equal to a given fraction of the asymptotic capacity $C$ of the respective channel, $N$ must be proportional to $V/C^2$, where $V$ is the respective channel dispersion quantity.

\end{abstract}


%
\IEEEpeerreviewmaketitle

\section{Background and Motivation}

{\em How many classical bits per channel use can be reliably communicated over a quantum channel?} This has been a central question in quantum information theory in an effort to understand the intrinsic limit on the classical capacity of physical quantum channels such as optical fiber or free-space optical channels. The Holevo limit of a quantum channel is an upper bound to the Shannon capacity of the classical channel {\em induced} by pairing the quantum channel with any specific transmitted states, modulation format, and the choice of a receiver measurement~\cite{PhysRevA.54.1869, Holevo98}. The Holevo limit is in principle also an achievable information rate, and is known for several important practical channels, such as the lossy-noisy bosonic channel~\cite{giovannetti2004classical, patron2014}. However, a receiver that attains the Holevo capacity, must in general make joint ({\em collective}) measurements over long codeword blocks. Such measurements cannot be realized by detecting single modulation symbols followed by classical post processing. 

The phenomenon that a joint-detection receiver (JDR) is able to yield a higher information rate (in error-free bits communicated per use of the quantum channel) than what is possible by any single-symbol receiver measurement is known as {\em superadditivity} in the classical capacity of a quantum channel~\cite{sasaki1997demonstration,Sas98}. We would like to clarify that the more prevalent use of the term {\em superadditivity} of capacity refers to the scenario when a quantum channel has a higher classical communication capacity when using transmitted states that are entangled over multiple channel uses~\cite{Has09}. For the bosonic channel, it was shown that entangled inputs at the transmitter cannot get a higher capacity~\cite{giovannetti2004classical}. However, one {\em can} get a higher capacity on the bosonic channel---as compared to what is possible by any optical receiver that measures one channel output at a time---by using joint-detection measurements at the receiver. As the number of symbols over which the receiver jointly acts increases, the capacity steadily increases. In this paper, we use the term {\em superadditivity} in this latter context, and provide a general lower bound on the scaling of the capacity with the length of the joint measurement. This usage of the term was first adopted by Sasaki {\it et al.}~\cite{sasaki1997demonstration}, and the phenomenon of superadditivity was demonstrated in~\cite{sasaki1997demonstration,Sas98,bennett1997entanglement}, for example, by showing a gap between the Holevo capacity and the maximum information rate achievable with the optimal single-symbol receiver measurement.

There are several JDR measurements that are known to achieve the Holevo capacity---the square-root measurement (SRM)~\cite{PhysRevA.54.1869}, the Yuen-Kennedy-Lax (YKL) minimum probability of error measurement~\cite{Hel76, yuen1975optimum}, the sequential-decoding measurement~\cite{Llo10, giovannetti2012achieving, Wil12,sen2012achieving}, the successive-cancellation decoder for the quantum polar code~\cite{Wil12a, Wil12b}, a two-stage near-unambiguous-detection receiver~\cite{Tak13}, and a bisection-decoding protocol~\cite{gio16}. There are a few characteristics that are common to each one of these measurements. First, the size of the joint-detection measurement is tied to the blocklength of the code, i.e., the measurement must act on the entire codeword and hence its size must increase with the length of the codeword. Second, none of these measurement specifications translate readily into a realizable receiver in the context of optical communication. Since it is known that a simple laser-light (coherent-state) modulation achieves the Holevo capacity of the lossy bosonic channel~\cite{giovannetti2004classical}, almost all the complexity in achieving the ultimate limit to the reliable communication rate lies at the receiver. Finally, none of these capacity-achieving measurements tell us how the achievable information rate increases exclusively with the size of the receiver measurement (while imposing no constraint whatsoever on the classical coding complexity). 

The complexity of implementing a joint quantum measurement over $N$ channel symbols in general grows exponentially with $N$~\cite{solovaykitaev}. This is because a general length-$N$ projective measurement can always be realized by a (quantum) unitary transformation on the $N$ channel outputs followed by product single-symbol measurements on each output of the unitary. Even though it is possible that this worst-case exponential scaling of resources with $N$ may be averted for codes with specific symmetries~\cite{kroviguha}, the sheer physical complexity of realizing joint operations---which would involve highly non-classical transformations of the received optical field within an optical receiver---, and the fact that no JDR realization that can even in principle outperform conventional optical receivers exists, make it of great practical interest to find how the maximum achievable information rate (error-free bits per channel use) scales with the size of the JDR.
 
In this paper, we shed some light on this problem, for classical communication over a quantum channel whose outputs are pure quantum states---the so-called pure-state classical input-quantum output (cq) channel. The lossy bosonic channel is an important practical example of such a channel, since transmitting a coherent state (the quantum description of ideal laser light), which is a pure state, results in an attenuated (pure) coherent state at the channel output. We prove a general lower bound on the finite-measurement-length capacity of a pure-state cq channel.

Finally, we would like to remark on an important difference between our setup in this paper and the setups in~\cite{Tom13, Mat12, wilde2014second} to find the finite-blocklength rate over a cq-channel and second-order asymptotics (channel dispersion). 
The latter papers explore how the achievable quantum channel coding rate $(\log_2 M_{N,\epsilon})/N$ (bits per channel use), at a {\em given decoding error threshold} $\epsilon$, increases when both the code length and the measurement length increase {\em together}\footnote{$M_{N,\epsilon}$ is the maximum number of messages that can be transmitted over a finite number ($N$) of uses of the quantum channel with average error probability $\epsilon$ (with no further outer coding permitted). $\log_2M_{N,\epsilon}$ can also be thought of as the one-shot classical capacity of the $N$-fold tensor product quantum channel (the amount of classical information that can be transmitted through a single use of the tensor product channel) such that the error probability is below $\epsilon$~\cite{wangrenner2012}.}. We consider the {\em asymptotic} capacity $C_N/N$ (error-free bits per channel use)~\cite{holevo98arxiv}, while imposing a constraint on the receiver to make collective measurements over $N$ channel outputs, but with no restriction on the complexity of any classical outer code that may be used on the classical channel induced by the $N$-length joint quantum measurement. In other words, we impose the length restriction only on the front end of receiver that acts on the quantum channel outputs (e.g., an $N$-symbol block of optically-modulated laser pulses received at the receiver) thereby producing a classical output (e.g., an electrical photocurrent), which in turn could be post-processed by an arbitrarily complex, including soft-information-processing, classical algorithm.

\section{Review of Concatenated Coding and Problem Statement}\label{sec:Int_0}

In order to separate the complexity of the quantum receiver and the complexity of the classical decoder, it is natural for us to consider a {\it concatenated coding} scheme over a quantum channel, which we describe below.
The idea of concatenated coding, which is comprised of an inner code and an outer code, was first introduced by Forney in~\cite{forney1966concatenated}, as a coding scheme for classical channels to reduce classical decoding complexity while achieving reliable communication up to the Shannon capacity $C^{(c)}$ of a classical channel. In Section~\ref{sec:Int_A}, we start by reviewing Forney's concatenated coding concept over a classical channel. We then show in Section~\ref{sec:Int_B} how this structure can be adopted for quantum channels to separate the complexity of the quantum receiver and that of the classical post processing. Under the concatenated coding structure, the quantity $C_N/N$ is defined as the maximum achievable information rate with $N$ being the length of an inner code on which the quantum (joint) receiver acts, while the length of the outer code may go to infinity. In Section~\ref{sec:Int_C}, we revisit concatenated coding over a classical DMC and pose the question of how the maximum achievable (error-free) information rate increases with the size $N$ of the inner-code blocklength, in the asymptotic limit of the length of the overall codeword going to infinity. This setup is different compared to Forney's original analysis where both the inner code and the outer-code blocklengths are assumed to increase without bound. 
We define the quantity $C^{(c)}_N/N$ as the maximum information rate transmissible over a classical channel with a concatenated coding scheme under the constraints that the number of the inner-code messages does not exceed $e^{NC^{(c)}}$ and the inner decoder is forced to make hard decisions on $N$-length inner codewords.
In Section~\ref{sec6}, we discuss the operational meanings of both of the aforesaid quantities ($C_N/N$ for a pure-state cq-channel, and $C^{(c)}_N/N$ for a classical DMC), and develop a unifying framework to address their properties. We explain the organization of the rest of this paper in Section~\ref{sec:Int_D}.

Before we proceed, let us recall that all classical DMC models employed in optical (or microwave) frequency communications are obtained by first starting from an underlying physical (quantum) channel, then picking a specific product-state modulation format which induces a cq channel, and finally picking a receiver measurement that detects one quantum channel output (i.e., optically-modulated pulses) at a time, producing an electrical output, which in turn induces a classical DMC. Most of the `noise' in the DMC results from the last step, that of the receiver converting the modulated symbol in the electromagnetic domain to the electrical (classical) domain. Because of the progression of restrictions employed above, it is clear that the regularized Holevo capacity of the underlying physical quantum channel is in general greater than the Holevo capacity of the induced cq channel, which in turn is in general greater than the capacity of the induced classical DMC. The Shannon capacity $C^{(c)}$ of this DMC should therefore satisfy: $C^{(c)}_N/N \le C^{(c)} \le C_1 \le C_N/N \le C$, where $C$ is the Holevo capacity of the cq channel. However, the mathematical similarity of $C^{(c)}_N/N$ and $C_N/N$, and their unifying treatment we develop in this paper, need not be tied to the above context of a DMC being induced by a cq channel paired with a single-symbol receiver. Our unified treatment may help translate any future development in quantifying the superaddivity---potentially better than what we do in this paper---for one problem, to the other; both problems being practically important in their own right.

\subsection{Concatenated Coding over a Classical Channel}\label{sec:Int_A}
Let us first review a block coding scheme for a classical DMC. 
For a classical channel with the transition probability $P_{Y|X}(y|x)$ with inputs $x\in\mathcal{X}$ and outputs $y\in\mathcal{Y}$, a block coding scheme  for $N$ uses of the channel is specified by an encoder and a decoder.
The encoder maps some finite message set  $\mathcal{J}$ into length-$N$ sequences in $\mathcal{X}^N$.  The elements of $\mathcal{J}$ are called messages, and the images of the messages under the encoding map are called codewords. The rate $R$ of such a code is  $R=\frac{1}{N}\log|\mathcal{J}|$. 
When one of the messages, $j\in\mathcal{J}$, is chosen, the length-$N$ codeword $\mathbf{x}_j=(x_{j,1},\dots,x_{j,N})\in\mathcal{X}^N$ is transmitted by  $N$ uses of the channel.
The probability that a length-$N$ output sequence $\mathbf{y}=(y_1,\dots,y_N)\in\mathcal{Y}^N$ is observed is characterized by the product of the transition probabilities, $P_{Y^N|X^N}(\mathbf{y}|\mathbf{x}_j):=\prod_{i=1}^NP_{Y|X}(y_i|x_{j,i})$. Given the channel outputs, the decoder processes the output sequence and maps it into an element in a set $\mathcal{K}$. Commonly the observer wants to know which message was transmitted. For such a case, ${\mathcal{K}}=\mathcal{J}$ and the decoder outputs an estimate $k\in\mathcal{J}$ of the transmitted message. When this decoding map is deterministic, we can define the set of the length-$N$ output sequences $\mathbf{y}\in\mathcal{Y}^N$ that are decoded to the message $j$, and denote the set as $\mathcal{Y}_j\subset\mathcal{Y}^N$. 
The probability of decoding error depends on the encoder, channel, and decoder. 
If it is assumed that all messages are equally likely, the average decoding error probability is defined as 
\beq\label{eqn:pe_inner_kj}
p_e=e^{-NR}\sum_{j=1}^{e^{NR}}\sum_{\mathbf{y}\notin \mathcal{Y}_j}P_{Y^N|X^N}(\mathbf{y}|\mathbf{x_j}). 
\eeq

Once the block coding scheme  for a DMC is specified, for each message $j\in\mathcal{J}$, the probability that the decoder outputs an estimate $k\in\mathcal{K}$ is specified.  
For a good block coding scheme with $N$ large, for each message $j$ the estimate $k$ would generally match the transmitted message. But for a finite $N$, the error probability $p_e$ may not be close to $0$. 
The encoder, $N$ uses of the DMC, and  decoder, collectively form a discrete memoryless {\it superchannel}, whose transition probabilities are given by: $p^{(N)}_{k|j}:=\sum_{\mathbf{y}\in \mathcal{Y}_k}P_{Y^N|X^N}(\mathbf{y}|\mathbf{x_j})$ for $j\in\mathcal{J}$ and $k\in\mathcal{K}$. 

\begin{figure}[t]
\centerline{\includegraphics[width=\columnwidth]{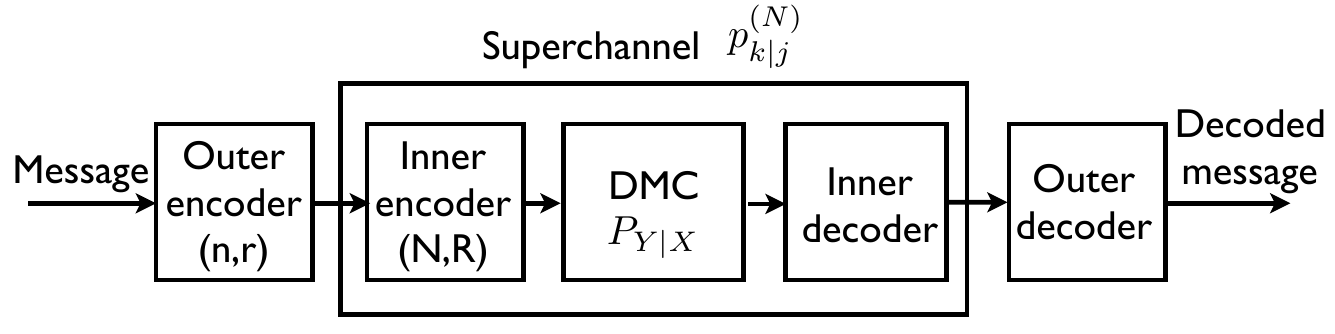}}
\caption{Concatenated coding over a classical DMC.}
\label{fig:classic1}
\end{figure}

For such a superchannel, it is possible to design another layer of block code of length $n$ and rate $r$. Such a code over the superchannel is called the {\em outer code}, whereas the aforementioned block code that forms the superchannel is called the {\em inner code}.
An outer encoder maps each message $m\in\{1,\dots, e^{nr}\}$ into a length-$n$ outer codeword $\mathbf{j}_m=(j_{m,1},j_{m,2},\dots,j_{m,n})\in \mathcal{J}^n$ where $\mathcal{J}\in \{1,\dots,e^{NR}\}$. Each symbol $j_{m,i}\in \mathcal{J}$, $i=1,\dots, n$, can be regarded as an inner-code message, and it is mapped to a length-$N$ inner codeword $\mathbf{x}_{j_{m,i}}\in \mathcal{X}^N$ by an inner encoder of length $N$ and rate $R$.
Upon transmission of this codeword through $N$ uses of the channel, the inner decoder receives a length-$N$ output sequence, and after processing the sequence it provides an estimate $k_i\in\mathcal{K}$ of the inner-code message. After collecting $n$ outputs  $\mathbf{k}=(k_1,\dots, k_n)\in\mathcal{K}^n$ of the inner decoder, the outer decoder processes those to generate an estimate $\hat{m}\in\hat{\mathcal{M}}$ of the outer-code message $m$.
The total length of the concatenated code is thus $N_c=nN$ and the rate of the code is $R_c=\frac{1}{nN}\log{e^{nr}}=\frac{r}{N}$.
A concatenated code over a DMC $P_{Y|X}$ is illustrated in  Fig. \ref{fig:classic1}.

In~\cite{forney1966concatenated}, it was demonstrated that there exist concatenated codes that can achieve the capacity of the DMC with the decoding complexity increasing only linearly in a small power of the overall blocklength $N_c=nN$, when both the inner-code blocklength $N$ as well as the outer-code blocklength $n$ go to infinity. The idea is to use an inner code with the  maximum likelihood inner decoder, paired with an algebraic outer code that admits an efficient decoding algorithm whose complexity is proportional to a small power of the outer-code blocklength. Even when there exists a loss of information at the inner decoder due to a hard estimate of the inner-code message, by designing the inner code based on the channel coding scheme that achieves the capacity as $N\to\infty$, the loss of information does not hurt the overall rate of the concatenated codes in the asymptotic regimes of $N$ and $n$.
Moreover, even though the complexity of the maximum-likelihood inner decoder increases exponentially in the inner-code blocklength $N$,  by increasing $n$ and $N$ with a significantly different order, e.g., $n=e^{NR}$, the overall complexity of the decoding algorithms becomes proportional to a small power of  the overall blocklength $N_c=nN$.
Here the decoding complexity is measured by the increasing rate of the number of  classical computations such as a comparison of likelihoods, in terms of  the overall blocklength $N_c$. In the next section, we show how this concatenated coding scheme can be adopted to analyze the maximum achievable information rate over a quantum channel, while separating the complexity of the quantum detection and that of classical processing.

\subsection{Concatenated Coding over a Classical Input-Quantum Output Channel and Superadditivity of $C_N$}\label{sec:Int_B}

\begin{figure}[t]
\centerline{\includegraphics[width=\columnwidth]{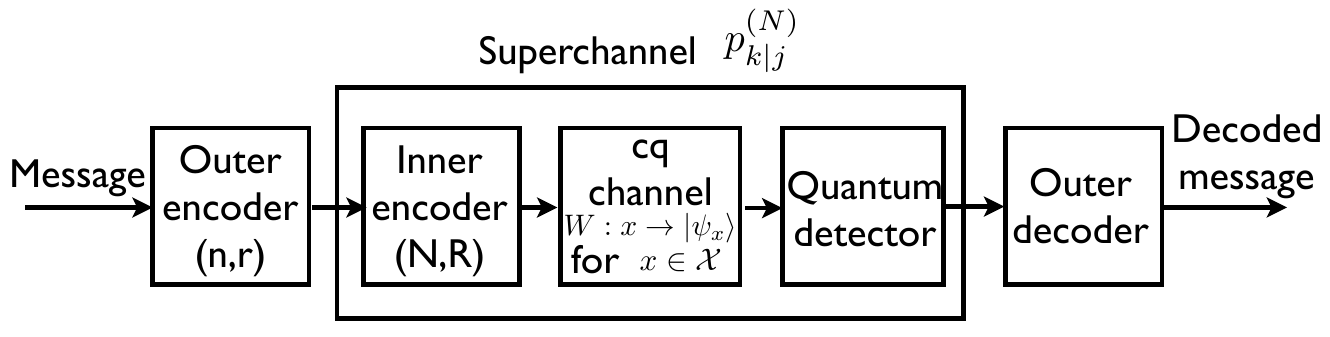}}
\caption{Concatenated coding over a pure-state classical input-quantum output (cq) channel.}
\label{fig:quantum1}
\end{figure}

In a future optical communication system that can employ joint detection and achieve an information rate higher than the Shannon capacities associated with any of the conventional optical receivers, the number $N$ of channel symbols jointly detected (using quantum-limited detection on an $N$-symbol block of received modulated pulses) will likely be much harder to scale up compared to the length of any classical code that may be employed. It is therefore hard to motivate analyzing quantum communication systems where both the blocklength of the channel code and the length of the joint-detection receiver increase {\em together} asymptotically. Instead, we consider the practically relevant model where the two infinities are decoupled via a concatenated coding scheme over a quantum channel, as depicted in Fig.~\ref{fig:quantum1}. 
In this model, the quantum joint-detection receiver acts on finite-blocklength (length $N$) modulated (quantum) inner codewords, while the overall codeword length $N_c=nN$ goes to infinity, as the classical outer-code length $n$ goes to infinity.

It is instructive to compare the concatenated coding scheme for the cq channel shown in Fig.~\ref{fig:quantum1} with the concatenated coding scheme over a DMC, shown in Fig. \ref{fig:classic1}. At the core of Fig.~\ref{fig:quantum1} is a cq channel $W:x\to\ket{\psi_x}$ for inputs $x\in\mathcal{X}$, and the box marked ``Quantum detector" is the inner decoder that acts on a product-state modulated inner codeword---a sequence of $N$ pure states each chosen from $\left\{|\psi_x\rangle\right\}$. 
Similar to the classical case, after an outer encoder maps each message $m\in\{1,\dots, e^{nr}\}$ into a length-$n$ outer codeword $\mathbf{j}_m=(j_{m,1},j_{m,2},\dots,j_{m,n})\in \mathcal{J}^n$ where $\mathcal{J}\in \{1,\dots,e^{NR}\}$,   each symbol $j_{m,i}\in \mathcal{J}$ is mapped to a length-$N$ inner codeword $\mathbf{x}_{j_{m,i}}\in \mathcal{X}^N$ by the inner encoder of length $N$ and rate $R$.
The length-$N$ inner codeword $\mathbf{x}_{j_{m,i}}$ is then mapped to a length-$N$ sequence of quantum states by $N$ uses of the quantum channel.
The quantum detector jointly measures each length-$N$ sequence of quantum states and generates a classical output $k_i\in\mathcal{K}$ for $i=1,\dots, n$ where $\mathcal{K}$ is the set of possible outcomes from the quantum detector.
In general, $ \mathcal{K} \supset\mathcal{J}= \{1,\dots, e^{NR}\} $.
When $\mathcal{K}=\mathcal{J}$, the quantum detector (inner decoder) basically gives an estimate of the encoded inner-code message $j_{m,i}$.
After the classical outer decoder collects $n$ outputs of the quantum detector $\mathbf{k}=(k_1,\dots, k_n)\in\mathcal{K}^n$, it processes the sequence of outputs to find an estimate of the transmitted message. 

The inner encoder, $N$ uses of the  cq channel, and the quantum joint-detection receiver, collectively form a discrete memoryless {\it superchannel}, with transition probabilities $p^{(N)}_{k|j}:=\Pr(K=k|J=j)$ where $J$ and $K$ indicate the input and the output random variables of the superchannel, respectively. 
Note that both the input and the output of this superchannel are classical, even though quantum operations happen inside the superchannel---which may comprise non-standard physical (e.g., optical) means to convert the quantum state of a length-$N$ inner codeword (e.g., a train of $N$ modulated laser-light pulses) to one classical output $k_i$, via a joint-detection method that is not even physically describable as a sequence of detections of each of the $N$ channel symbols followed by classical (such as soft-information) post processing.

We define the maximum mutual information of this superchannel, over all choices of inner encoders of blocklength $N$ and rate $0\leq R\leq \log|\mathcal{X}|$, and over all choices of quantum detectors that jointly measure the length-$N$ output quantum states as:
\begin{equation}\label{eqn:CNdef_sec1}
C_N:=
\max_{\substack{p_j,\\j\in\{1,\dots,|\mathcal{X}|^{N}\}}} \max_{\Big\{\substack{\text{$N$-symbol inner encoder-}\\\text{measurement pairs}}\Big\}}I(p_j,p_{k|j}^{(N)})
\end{equation}
where the mutual information $I(p_j,p_{k|j}^{(N)}):=\sum_{j\in\{1,\dots,|\mathcal{X}|^N\}} p_j\left(\sum_{k\in\mathcal{K}} p_{k|j}^{(N)}\log\frac{p_{k|j}^{(N)}}{\left(\sum_{{j'\in\{1,\dots,|\mathcal{X}|^N\}}}  p_j' p_{k|j'}^{(N)} \right)}\right)$ is evaluated with the input distribution $p_j:=\Pr(J=j)$, $j\in\{1,\dots, |\mathcal{X}|^N\}$, and the superchannel distribution $p_{k|j}^{(N)}=\Pr(K=k|J=j)$, $k\in\mathcal{K}$, which is determined by the $N$-symbol inner encoder-measurement pairs.

Note that, in general, in order to attain this maximum mutual information of the superchannel induced by the inner encoder-measurement pair, the number of outputs of the quantum detector, $|\mathcal{K}|$, may need to be greater than $|\mathcal{X}|^N$~\cite{Dav78, shor2004adaptive}. 
In fact, it is known that the number of outputs of the quantum detector that maximizes the mutual information for $M$ linearly independent pure states, grows as $O(M^2)$~\cite{Dav78}. In our case, $M=|\{j\in\{1,\dots,|\mathcal{X}|^N\}:p_j^*>0\}|$ for $p_j^*$ being the input distribution that maximizes the mutual information in~\eqref{eqn:CNdef_sec1}.
But by considering the case when the output of the inner decoder makes a hard decision on the inner-code message, we can find a lower bound on $C_N$, which gives practically important results as will be discussed later in this paper.

A classical channel coding scheme that achieves the maximum mutual information of a DMC can be used to design outer codes that reliably communicate information through this superchannel at a rate arbitrarily close to the maximum mutual information $C_N$ as the outer code length $n\to\infty$. By Shannon's coding theorem, for any rate $r<C_N$, there exists an outer code of length $n$ and rate $r$ that can be decoded by {\em classical} processing with arbitrarily small decoding error probability as $n\to \infty$.
Since the rate of the overall concatenated code is $R_c=r/N$, the maximum information rate achievable by the concatenated code {\it per use of the quantum channel} can approach $C_N/N$ for a finite $N$.

From the definition of $C_N$, superadditivity of the quantity, i.e., $C_{N_1}+C_{N_2}\leq C_{N_1+N_2}$, can be shown.  In~\cite{holevo1977problems}, Holevo showed that the limit $\lim_{N\to \infty}C_N/N$ exists and is equal to the ultimate capacity of the quantum channel, which is also equal to the Holevo capacity~\cite{Holevo98,schumacher1997sending}, as defined in~\eqref{eq:HolevoCapacity} in Section~\ref{sec:cq}.

The question we want to answer is: {\em How does the maximum achievable information rate $C_N/N$ change as the length of the quantum measurement, $N$, increases? Or more precisely, how does $\max_{m\in\{1,2,\dots,N\} }\{C_m/m\}$ increase with $N$?\footnote{The superadditivity of $C_N$, i.e.,  $C_{N_1}+C_{N_2}\leq C_{N_1+N_2}$, implies that for any positive integers $m$ and $k$ such that $m/k$ is an integer, $C_{m}$ is greater than or equal to $kC_{m/k}$, i.e., $C_{m}/(m)\geq C_{m/k}/(m/k)$. Therefore,  $\max_{m\in\{1,2,\dots,N\} }\{C_m/m\}$  can be simplified by removing the elements that are smaller than or equal to $C_m/m$, i.e., $\{C_{m/k}/(m/k): k\in\{1,\dots,m-1\}, m/k\in \mathbb{N}\}$, from the maximization, starting from $m=N$ and continuing in the decreasing order.  }} 

Since quantum processing occurs only at the inner decoder, the complexity of the quantum processing only depends on $N$, but not on the outer code length $n$. Therefore, the trade-off between the rate and the (quantum) complexity of the measurement device can be captured by how fast  $\max_{m\in\{1,2,\dots,N\} }\{C_m/m\}$ increases in $N$. It is known that for some examples of classical input-quantum output channels, strict superadditivity of $C_N$, i.e., $C_{N_1}+C_{N_2}< C_{N_1+N_2}$, holds~\cite{peres1991optimal,holevo98arxiv, Guh10}. However, the calculation of $C_N$, even for a pure-state binary alphabet, is extremely hard for $N>1$ because the complexity of the optimization increases exponentially with $N$. 

Instead of aiming to calculate $C_N$ exactly for a specific cq channel, in this paper, we derive a lower bound on $C_N/N$, which becomes tight for $N$ large enough. Using this bound, we will show that it is possible to calculate the inner-code blocklength $N$ at which a given fraction of the ultimate (Holevo) capacity is achievable. 

\subsection{Concatenated Codes over a Classical DMC with a Finite Blocklength Inner Code}\label{sec:Int_C}

In this section, we revisit concatenated coding over a {\it classical} channel and define the asymptotic capacity $C_N^{(c)}/N$ of a classical DMC---analogously to $C_N/N$ in~\eqref{eqn:CNdef_sec1} for a cq channel---with the condition now being that the inner decoder has to make hard decisions on the inner-code message encoded by length-$N$ (inner) codewords of rates $0\leq R\leq C^{(c)}$ where $C^{(c)}$ is the capacity of the DMC. In a later section of this paper, we will establish a lower bound on $C_N^{(c)}/N$, which can be treated in a unifying framework with the lower bound on $C_N/N$ of a classical-quantum channel. 

In~\cite{forney1966concatenated}, when Forney analyzed the performance of concatenated codes, he first investigated the coding-theoretical limits on the attainable mutual information of superchannels induced by an inner code of length $N$ and rate $0\leq R\leq C^{(c)}$ under different assumptions on the inner decoder outputs, without any restrictions on the complexity of outer codes. It is obvious that when the inner decoder generates a sufficient statistic of the channel output and forwards it to the outer decoder, there is no loss of information, so that the performance of the concatenated code can be as good as an optimal code, even with the restricted structure of code concatenation. Despite the fact that the performance remains intact, for this case, the decoding complexity increases exponentially with the overall length of the code. 
On the other hand, even if there is some loss of information at the inner decoder by making a hard decision on the message of the inner code, as the inner-code blocklength $N$ as well as the outer-code blocklength $n$ go to infinity, the capacity of the underlying classical DMC can be achieved with the concatenated code. This was demonstrated in~\cite{forney1966concatenated} by analyzing a lower bound on the maximum mutual information of superchannels  where the inner decoder makes a hard decision on the transmitted inner codeword. As $N\to\infty$, the maximum mutual information of the superchannel approaches $NC^{(c)}$ where $C^{(c)}$ is the capacity of the DMC. Moreover, the average decoding error probability $p_e$ of the inner code, defined in~\eqref{eqn:pe_inner_kj}, decreases exponentially in $N$, for any rate $R<C^{(c)}$ of the inner code.
These properties of the superchannel were essential in achieving the capacity of a DMC using concatenated coding, even after the outer code was restricted to algebraic codes that enable a simple decoding algorithm. 

Even though concatenated coding allows reliable communications up to the capacity of a DMC with a much improved decoding complexity, this coding scheme results in a great delay to decode the message, since the outer-code blocklength $n$ increases on the order of $e^{NR}$ with the inner-code blocklength $N\to\infty$. The delay issue is critical in modern data communications, especially in wireless communications. Therefore, it is of great interest to study trade-offs between information rate and delay in communication systems.

 We study one aspect of this trade-off  by asking the following question for the concatenated code over a classical DMC, similar to the one we asked for the quantum channel but with additional assumptions: Assume that the cardinality of the set of inner-code messages at a finite blocklength $N$ does not exceed $\lfloor e^{NC^{(c)}} \rfloor$ and the inner decoder makes a hard estimate on the messages of the inner code {\it at a finite blocklength $N$}. Under these conditions, what is the asymptotic capacity (error-free bits per use of the underlying classical DMC) at a fixed inner-code blocklength $N$ as the outer-code blocklength $n\to\infty$ (i.e., with no restriction on the complexity of the outer code)? 
 When the inner decoder makes a hard estimate for the inner-code message encoded in length-$N$ inner codewords, even though the estimate may contain errors, it still allows an observer to receive a rough estimate of the inner-code message after every $N$ transmissions. 
 After the outer decoder collects a length-$n$ output sequence of the hard-decision inner decoder and processes the output sequence to decode the outer-code message, the erroneous information can be corrected and the outer-code message can also be reliably decoded.

With the inner decoder that makes a hard decision at a finite blocklength $N$ for the inner code of rate $R\leq C^{(c)}$, the maximum achievable information rate by the concatenated code is defined to be $C^{(c)}_N/N$ where,
\begin{equation}\label{eqn:classCNhard1}
C^{(c)}_N=\max_{\substack{p_j,\\ j\in\{1,\dots, \lfloor e^{NC^{(c)}} \rfloor\}}}\max_{\Big\{\substack{\text{N-symbol inner encoder-}\\ \text{{\em hard-decision} decoder pairs}}\Big\}}I(p_j,p_{k|j}^{(N)})
\end{equation}
for the input distribution $p_j$, $j\in\{1,\dots, \lfloor e^{NC^{(c)}}\rfloor\}$, and the superchannel distribution $p_{k|j}^{(N)}$ between the inner-code message $j$ and hard-decision inner decoder output $k$. 
Let $M$ denote the cardinality of the set of inner-code messages with a positive probability, i.e., $M=|\{j\in\{1,\dots, \lfloor e^{NC^{(c)}} \rfloor\}: p_j>0\}|$. 
In the maximization  on the right hand side of~\eqref{eqn:classCNhard1}, we consider only an inner encoder with messages of cardinality $M\leq \lfloor e^{NC^{(c)}} \rfloor$ and a decoder whose output set has the same cardinality as that of the inner-code messages.
Note that when we defined $C_N$ in~\eqref{eqn:CNdef_sec1} for a quantum channel we imposed the similar length-$N$ constraints on the inner codeword and on the quantum detector but did not restrict the rate of the inner code nor the cardinality of the outputs of the quantum detector. These additional constraints in the definition of $C_N^{(c)}$ are to develop mathematical analogy between $C_N$ and $C_N^{(c)}$ based on the fact that the length-$N$ quantum detector in the quantum case is analogous to the inner decoder making hard decisions on length-$N$ inner codewords in the classical case, which will be discussed in more details in Section~\ref{sec6}. We will use this analogy to provide a general lower bound on $C_N/N$.

Without the two additional constraints on the cardinality of the inputs and outputs of the superchannel in the definition of $C_N^{(c)}$ for a classical DMC, one can easily define an inner encoder that is a trivial one-to-one mapping from the message set $\mathcal{J}=\{1,\dots, |\mathcal{X}|^N\}$ of rate $R=\log|\mathcal{X}|\geq C^{(c)}$ to $\mathcal{X}^N$  and an inner decoder that is also a one-to-one mapping from $\mathcal{Y}^N$ to $\mathcal{K}=\{1,\dots, |\mathcal{Y}|^N\}$, and show that
\beq
C^{(c)}_N=\max_{P_{X^N}}I(X^N;Y^N)=NC^{(c)}
\eeq
for any finite $N$. But this $C^{(c)}_N$ does not say how much information about the inner-code message can be extracted by the inner decoder at a finite blocklegnth $N$, since the probability of getting a correct estimate of the inner-code message can never converge to 0 even when $N\to\infty$.
Therefore, to exclude the trivial encoders and decoders of one-to-one mappings and to justify the operational meaning of $C^{(c)}_N$, we restrict ourselves to look at only an inner code of rate $0\leq R\leq C^{(c)}$ with a hard decision decoder, along with the definition of $C_N^{(c)}$ as in~\eqref{eqn:classCNhard1}.

Moreover, these constraints on the cardinality of the inner encoder and decoder make $C^{(c)}_N$ exhibit non-trivial superadditivity.
From the definition of $C_N^{(c)}$, superadditivity of the quantity, i.e., $C_{N_1}^{(c)}+C_{N_2}^{(c)}\leq C^{(c)}_{N_1+N_2}$, can be shown. 
Moreover, Shannon's coding theorem implies that the limit of $C_N^{(c)}$ is $
\lim_{N{\to\infty}}C_{N}^{(c)}=NC^{(c)}
$ where $C^{(c)}$ is  the Shannon capacity of a classical DMC.
For a finite $N$, on the other hand, the quantity $C_{N}^{(c)}$ might be strictly smaller than $NC^{(c)}$.
For example, consider a binary symmetric channel BSC$(\delta)$ with a flipping probability $\delta=0.1$. The Shannon capacity of the BSC($\delta$) is $C^{(c)}=\log2+(\delta\log\delta+(1-\delta)\log(1-\delta))=0.368$ nats/channel use.
At $N=1$, we have $C_{1}^{(c)}=0$ since the maximum cardinality of the input set that satisfies the condition $M\leq \lfloor e^{C^{(c)}}\rfloor$ is only 1. 
At $N=2$, the maximum cardinality of the input set is 2. By having $\{00,11\}$ as codewords for the two messages and by using the hard-decision decoder that maps the outputs $\{00,01, 10\}$ to the input message $\{00\}$ and the output $\{11\}$ to the input message $\{11\}$ we can achieve the maximum information rate of $C_2^{(c)}/2\approx 0.205$ at the optimal input distribution $p_1\approx 0.55$, $p_2=1-p_1$. Therefore, for a BSC(0.1), we demonstrated the strict superadditivity of $C_N^{(c)}$ by showing $C^{(c)}_1<C_2^{(c)}/2<\lim_{N\to\infty}C_N^{(c)}/N=C^{(c)}$. 

We are interested in how $\max_{m\in\{1,2,\dots N\}}\{C^{(c)}_{m}/m\}$ increases in $N$ for a classical DMC.
In this paper, we provide a lower bound on $C_N^{(c)}/N$ for a fixed $N$ and show a close mathematical connection between this  bound and the lower bound on $C_N/N$ of a cq channel. 
\subsection{Organization of the Paper}\label{sec:Int_D}

The rest of this paper is organized as follows. In Section~\ref{sec:cq}, we introduce the notation, some fundamentals of classical input-quantum output (cq) channels, and the Holevo capacity of these channels. In Section \ref{sec3}, examples of quantum channels where strict superadditivity $C_1<C$ holds, will be demonstrated. Our main theorem, which states a lower bound on $C_N/N$, which strictly increases with $N$, will be stated in Section \ref{sec4} with examples to show how to use the theorem to calculate a blocklength $N$ to achieve a given constant fraction of the Holevo capacity. This theorem will be proved in Section \ref{sec5}. Thereafter in Section \ref{sec6}, we will provide a lower bound on $C_N^{(c)}/N$ for a classical DMC and will compare it with the lower bound on $C_N/N$ of a cq channel under a unifying framework.
An approximation of the lower bounds on $C_N/N$ and on $C_N^{(c)}/N$ will also be provided by introducing quantum and classical versions of {\it channel dispersion} $V$.  Some future directions on the study of strict superadditivity will be discussed in Section \ref{sec:6con}.

\section{Classical Input-Quantum Output Channel}\label{sec:cq}

The classical capacity of a quantum channel is defined as the maximum number of information bits that can be sent per use of the quantum channel, by encoding a message into a transmitted modulated quantum state (which could be entangled over many uses of the channel), and decoding the message at the channel's output by applying any measurement permissible by quantum mechanics. In this paper, we will restrict our attention to a pure-state memoryless classical-quantum (cq) channel $W: x \to \ket{\psi_x}$, which takes a classical input $x\in\mathcal{X}$  at the input of the channel and maps it to a quantum (pure) state $|\psi_x\rangle \in \Hil$, where $\Hil$ is a complex Hilbert space.

As a concrete example of a pure-state cq channel, the transmission of an ideal laser light pulse over a lossy optical channel can be modeled as a pure-state cq channel ${\cal N}_{\eta}: {\pmb{\alpha}} \to |\sqrt{\eta}{\pmb{\alpha}}\rangle$, where ${\pmb{\alpha}} \in {\mathbb C}$ is the complex field amplitude (of the coherent state $|{\pmb{\alpha}}\rangle$) at the input of the channel, $\eta \in (0,1]$ is the transmissivity (the fraction of input power that appears at the output), and $|\sqrt{\eta}{\pmb{\alpha}}\rangle$ is a coherent state at the channel's output.\footnote{It is important to note here the difference between a classical channel and a classical-quantum channel. There is no physical measurement that can noiselessly measure the output amplitude $\sqrt{\eta}{\pmb{\alpha}}$. Any specific choice of an optical receiver---such as homodyne, heterodyne or direct-detection---induces a specific discrete memoryless {\em classical} channel $p({\pmb{\beta}} | {\pmb{\alpha}})$ between the input ${\pmb{\alpha}}$ and the measurement result ${\pmb{\beta}}$. The Shannon capacity of this induced classical channel, for any given measurement choice at the receiver, is strictly smaller than the Holevo capacity of the measurement-unrestricted cq channel ${\cal N}_\eta$. As discussed above, for a receiver's performance to asymptotically approach the Holevo capacity, it must make a joint-detection measurement over an infinite codeword, which in this case is a block of modulated laser pulses.}
  The coherent state $\ket{{\pmb{\alpha}}}$ is the quantum description of an ideal laser-light pulse in a given field mode, of mean photon number $|{\pmb{\alpha}}|^2$, and a carrier-phase offset given by the phase of ${\pmb{\alpha}}$. The coherent state is a pure state given by $\ket{{\pmb{\alpha}}} = e^{-|{\pmb{\alpha}}|^2/2}\sum_{n=0}^\infty \left({{\pmb{\alpha}}}^n/\sqrt{n!}\right)|n\rangle$, where $\ket{n}, n = 0, 1, \ldots$, the photon number states, form a complete orthonormal basis for the state space of a single optical mode.

In~\cite{PhysRevA.54.1869}, it was shown that the classical capacity of a pure-state cq channel $W$ is given by
\begin{equation}
C=\max_{P_X}\Tr(-\rho\log\rho),
\label{eq:HolevoCapacity}
\end{equation}
where $\rho=\sum_{x\in \cal{X}} P_X(x)\ket{\psi_x}\bra{\psi_x}$. This capacity is called Holevo capacity. The states $\ket{\psi_x}$, $x\in \cal{X}$, are normalized vectors in a complex Hilbert space $\Hil$, $\bra{\psi_x}$ is the Hermitian conjugate vector of $\ket{\psi_x}$, and $\rho$ is a {\it density operator}, a linear combination of the outer products $\ket{\psi_x}\bra{\psi_x}$ with weights $P_X(x)$. The Holevo capacity can also be written as $C=\max_{P_X}S(\rho)$, where $S(\rho)=\Tr(-\rho\log\rho)$ is the von Neumann entropy of the density operator $\rho$.

A length-$N_c$ block code for $N_c$ uses of a pure-state cq channel $W$ with input set $\mathcal{X}$ and output set $\{\ket{\psi_x}\} $ for $x\in \mathcal{X}$ consists of an encoder and a quantum measurement.
The encoder maps some finite message set $\mathcal{M}$ into elements of $\mathcal{X}^{N_c}$. 
For an input codeword $(x_1,\cdots,x_{N_c})\in\mathcal{X}^{N_c}$, the sequence of outputs of the quantum channel $W$ can be written as a tensor product state,  $\ket{\Psi_{x_1,\dots,x_{N_c}}}:=\ket{\psi_{x_1}}\otimes\cdots\otimes \ket{\psi_{x_{N_c}}}\in \{\ket{\psi_x}\}^{\otimes N_c}\subset \Hil^{\otimes N_c}$. When the received codeword is measured by an orthogonal projective measurement, $\{\ket{\Phi_k}\}$, $k\in\cal{K}$, which resolves the identity, i.e., $\sum_k \ket{\Phi_k} \bra{\Phi_k} =\mathbbm{1}$, in $\Hil^{\otimes N_c}$, the classical output $k$ is observed with probability equal to $|\braket{\Phi_k}{{\Psi_{x_1,\dots,x_{N_c}}}}|^2$, the magnitude squared of the inner product between the received codeword state and the measurement vector corresponding to the output $k$.  
When the received quantum codeword is measured by a more general Positive Operator Valued Measure (POVM) measurement, $\{\Pi_k\}$ such that $\Pi_k\geq 0$ and $\sum_{k}\Pi_k=\mathbbm{1}$ in $\Hil^{\otimes N_c}$, we observe the classical output $k$ with probability equal to $\bra{\Psi_{x_1,\dots,x_{N_c}}}\Pi_k\ket{\Psi_{x_1,\dots,x_{N_c}}}$.
The quantum measurement is designed to decode the received codewords with as small an error probability as possible. 
For any rate $R<C$, there exists a block code of length $N_c$ and rate $R$ that can be decoded with an arbitrarily small average probability of error as $N_c \to \infty$ by an appropriate quantum measurement acting jointly on the received codeword in $\Hil^{\otimes N_c}$~\cite{PhysRevA.54.1869, Holevo98}.

To achieve this capacity, however, a joint-detection receiver (JDR) needs to be implemented, which can measure the length-$N_c$ sequence of states jointly and decode it reliably among $e^{N_c R}$ possible messages. The number of measurement outcomes thus scales exponentially with the length of the codeword $N_c$, and the complexity of physical implementation (in terms of number of elementary finite-length quantum operations) of the receiver in general also grows exponentially with $N_c$. 

Considering this growth in complexity, in Section~\ref{sec:Int_B}, we proposed to limit the maximum length $N \le N_c$ of the sequence of states to be jointly detected at the receiver, independent of the length of the codeword $N_c$, and to analyze the asymptotic capacity $C_N/N$ in~\eqref{eqn:CNdef_sec1} at a fixed length $N$ of joint measurement. 
When we denote $p_{x^N}:=\Pr(X^N=x^N)$ as an input distribution over all possible length-$N$ inner codewords $x^N=(x_1,\dots,x_N)\in\mathcal{X}^N$ and denote the most general form of a length-$N$ quantum measurement (POVM) in $\Hil^{\otimes N}$ as $\{\Pi_k\}$ for $k\in\mathcal{K}$, $C_N$ defined in \eqref{eqn:CNdef_sec1} is equivalent to
\beq
C_N=\max_{p_{x^N}}\max_{\{\Pi_k\}}I(p_{x^N},p_{k|x^N})
\eeq
where $p_{k|x^N}:={\sf{Pr}}(K=k|X^N=x^N)=\bra{\Psi_{x_1,\cdots,x_N}}\Pi_k\ket{\Psi_{x_1,\cdots,x_N}}$. This form of definition for $C_N$ was first provided by Holevo in~\cite{holevo98arxiv}.

It is known that quantum measurements of fixed blocklengths cannot achieve the capacity of the quantum channel. Moreover, it was shown that for some examples of quantum channels, as the number $N$ of channel outputs jointly measured increases, the maximum number of information bits extracted per use of the quantum channel increases~\cite{peres1991optimal, sasaki1997demonstration}. In the next section, we show examples of quantum channels where strict superadditivity  can be demonstrated by showing that $C_1<C$.

\section{Strict Superadditivity of $C_N$}\label{sec3}

Before investigating how $\max_{m\in\{1,2,\dots,N\} }\{C_m/m\}$  increases with $N$ for a classical input-quantum output channel, we will show examples where strict superadditivity of $C_N$ can be shown by $C_1<C$ where $C_1$ is the maximum achievable information rate with symbol-by-symbol detection and $C$ is the ultimate capacity with quantum joint-detection receiver over infinite blocklength codewords. As discussed before, given a set of output quantum states $\{\ket{\psi_x}\}$, $x \in {\mathcal X}$, $C$ can be calculated from Holevo's result by finding the optimal input distribution that maximizes the von Neumann entropy $S(\rho)=\Tr(-\rho\log\rho)$ where $\rho=\sum_{x}P_X(x)\ket{\psi_x}\bra{\psi_x}$. Calculating $C_1$, on the other hand, requires finding a set of measurements as well as an input distribution to maximize the resulting mutual information, where the measurement acts on one channel symbol at a time. For general output quantum states, this is a hard optimization problem, since the measurement that maximizes $C_1$ may not be a projective measurement, and could be a Positive Operator Valued Measure (POVM)---the most general description of a quantum measurement---and furthermore the optimal POVM could have up to $|{\mathcal X}|\left(|{\mathcal X}|+1\right)/2$ outcomes~\cite{Dav78, shor2004adaptive}. 

However, for binary pure states $\{\ket{\psi_0},\ket{\psi_1}\}$, as shown in~\cite{holevo98arxiv}, $C_1$ and $C$ can be calculated as simple functions of the inner product $\gamma = |\langle \psi_0 | \psi_1 \rangle|$, and strict superadditivity can be shown. We will summarize this result of strict superadditivity in binary pure-state classical input-quantum output (cq) channels in Section~\ref{subsec:superadd_binary}.

We also consider the strict superadditivity of pure-state cq channels {\it with an input constraint}, in the context of optical communication, in Section~\ref{subsec:superadd_input}. 
The constraint will be the average energy of quantum states. A {\it coherent state} $\ket{\pmb{\alpha}}$ is the quantum description of a single spatio-temporal-polarization mode of a classical optical-frequency electromagnetic (ideal laser-light) field, where $\pmb{\alpha}\in \mathbb{C}$ is the complex amplitude, and $|\pmb{\alpha}|^2$ is the mean photon number of the mode. Since the energy of {\it a photon} with angular frequency $\omega$ is $E=\hbar \omega$ where $\hbar$ is the reduced Planck constant, the average energy (in Joules) of the coherent state $\ket{\pmb\alpha}$  of center frequency $\omega$, is $\hbar |\pmb\alpha|^2 \omega$. Note that the mean photon number $|\pmb\alpha|^2$ is a dimensionless quantity. Therefore, for propagation at a fixed center frequency $\omega$, an average energy constraint on the quantum states (or equivalently, an average power constraint with a fixed time-slot width) can be represented as a constraint on the mean photon number per transmitted mode. For example, for continuous output quantum states $\{\ket{\pmb\alpha}\}$ where $\pmb\alpha\in \mathbb{C}$, an average energy constraint $\hbar\omega\E$ per transmitted mode can be expressed as a constraint on the prior distribution $p(\pmb\alpha)$, with
$
\int |\pmb\alpha|^2 p(\pmb\alpha)d\pmb\alpha\leq \E,
$
where $\E$ is the constraint on the mean photon number per mode.

The important question of how many bits can be reliably communicated per use (i.e., per transmitted mode) of a pure-loss optical channel of power transmissivity $\eta \in (0,1]$, under the constraint on the average photon number per transmitted mode $\E$, was answered in \cite{giovannetti2004classical}. It was also shown that product coherent-state inputs are sufficient to achieve the Holevo capacity of this quantum channel. Since a coherent state $|\pmb\alpha\rangle$ of mean photon number $\E = |{\pmb\alpha}|^2$ transforms into another coherent state $|\sqrt{\eta}\,{\pmb\alpha}\rangle$ of mean photon number $\eta\E$ over the lossy channel, we will henceforth, without loss of generality, subsume the channel loss in the energy constraint, and pretend that we have a lossless channel $(\eta = 1)$ with a mean photon-number constraint $\mathbb{E} [|\pmb\alpha|^2]\leq \E$ per mode (or per `channel use'). The capacity of this channel is given by~\cite{giovannetti2004classical}
\begin{equation}\label{eqn:capacity_energy1}
C(\E)=(1+\E)\log(1+\E)-\E\log\E \text{  [nats/mode]},
\end{equation}
and it is achievable with a coherent-state random code with the amplitude $\pmb\alpha$ chosen from a circulo-complex Gaussian distribution with variance $\E$, i.e., $p(\pmb\alpha)=\exp[-|\pmb\alpha|^2/\E]/{(\pi \E)}$. However, achieving this ultimate capacity requires a joint-detection receiver that can jointly measure the infinite-length quantum codeword. Therefore, to understand the trade-offs between the maximum achievable information rate and the complexity of joint-detection receiver with an input constraint, we again consider a concatenated coding scheme with a finite-length quantum joint measurement as in Fig.~\ref{fig:quantum1}. We define $C_{N}(\E)$ as the maximum achievable information rate with the optimal inner encoder and joint measurements of length-$N$ under the mean photon-number constraint of $\E$,
\begin{equation}
\begin{split}
&C_{N}(\E):=\\
&\max_{\{ p(\pmb\alpha^N):  \int(\sum_{i=1}^N|\pmb\alpha_i|^2) p(\pmb\alpha^N) d\pmb\alpha^N \leq N\cdot\E\}}\max_{\{\Pi_y\}}I(p({\pmb\alpha^N}),p({y|\pmb\alpha^N})),
\end{split}
\end{equation}
where $p(\pmb\alpha^N)$ is an input distribution over $\pmb\alpha^N=(\pmb\alpha_1,\dots,\pmb\alpha_N)\in\mathbb{C}^N$ and $p(y|\pmb\alpha^N)=\bra{\pmb\alpha^N}\Pi_y\ket{\pmb\alpha^N}$ for a POVM $\{\Pi_y\}$ in $\Hil^{\otimes N}$.
We are interested in how $\max_{m\in\{1,\dots,N\}}\{C_m(\E)/m\}$ increases with $N$. Even though $C_1(\E)$ cannot be explicitly calculated because of the similar difficulties as in calculating $C_1$ for a general set of quantum states, by restricting the cardinality of the  quantum states to be binary, or by further restricting it to be a specific binary constellation, we can observe the strict superadditivity of pure-state cq channels  under the mean photon-number constraint. 

In the analysis of the capacity under the mean photon-number constraint, we will use the $o(\cdot)$ and $O(\cdot)$ notations to describe the behavior of functions of the mean photon number $\E$ in the regime of $\E\to 0$. A function described as $o(f(\E))$ and that described as $O(f(\E))$  satisfies
\beq
\lim_{\E\to 0}\Bigg|\frac{o(f(\E))}{f(\E)}\Bigg|=0,\quad \limsup_{\E\to 0}\Bigg|\frac{O(f(\E))}{f(\E)}\Bigg|<\infty,
\eeq
respectively.

\subsection{Strict Superadditivity for Binary Pure-State Channels}\label{subsec:superadd_binary}
The first step to calculate $C$ for the binary pure-state cq channel $W:x\to \ket{\psi_x}$, $x\in\{0,1\}$, is to find the eigenvalues of $\rho$ under an input distribution $\{1-q,q\}$. For $\rho =(1-q)\ket{\psi_0}\bra{\psi_0}+q\ket{\psi_1}\bra{\psi_1}$, the eigenvectors of $\rho$ have a form of $\ket{\psi_0}+\beta \ket{\psi_1}$ with some $\beta$ that satisfies
\begin{equation}
\begin{split}
\rho(\ket{\psi_0}+\beta\ket{\psi_1})&=\sigma (\ket{\psi_0}+\beta\ket{\psi_1})
\end{split}
\end{equation}
with eigenvalues $\sigma$. By solving this equation, we obtain the two eigenvalues as:
\begin{equation}
\begin{split}\label{eqn:eigenvalues}
&\sigma_1=\frac{1}{2}\left(1-\sqrt{1-4q(1-q)(1-\gamma^2)}\right), \,{\rm and}\\
&\sigma_2=\frac{1}{2}\left(1+\sqrt{1-4q(1-q)(1-\gamma^2)}\right),
\end{split}
\end{equation}
where $\gamma=|\braket{\psi_0}{\psi_1}|$. The resulting von Neumann entropy is
\begin{equation}
S(\rho)=\Tr(-\rho\log\rho)=-\sigma_1\log{\sigma_1}-\sigma_2\log{\sigma_2}.
\end{equation}
From this equation, it can be shown that $S(\rho)$ for the binary inputs is maximized at $q=1/2$, and the resulting capacity of the binary cq channel is
\begin{equation}\label{eqn:cap_bin1_1}
C=\max_{P_X}S(\rho)=-\frac{1-\gamma}{2}\log\frac{1-\gamma}{2}-\frac{1+\gamma}{2}\log\frac{1+\gamma}{2}.
\end{equation}
For the binary channel, $C_1$ is attained by the equiprior input distribution and a binary-valued projective measurement in the span of $\left\{|\psi_0\rangle, |\psi_1\rangle\right\}$---the same measurement that minimizes the average error probability of discriminating between equally-likely states $|\psi_0\rangle$ and $|\psi_1\rangle$. The derivation of $C_1$ for the binary case can be found in \cite{holevo98arxiv}, and is given by:
\begin{equation}\label{eqn:C1_bin1}
\begin{split}
C_1&=\frac{1-\sqrt{1-\gamma^2}}{2}\log\left(1-\sqrt{1-\gamma^2}\right)\\
&\quad\;+\frac{1+\sqrt{1-\gamma^2}}{2}\log\left(1+\sqrt{1-\gamma^2}\right).
\end{split}
\end{equation}
The capacity $C$ is strictly greater than $C_1$ for all $0<\gamma<1$, which demonstrates the strict superadditivity of $C_N$ for all binary pure-state cq channels with any output quantum states $\{\ket{\psi_0},\ket{\psi_1}\}$.

\subsection{Strict Superadditivity for Binary Coherent-State Channels with a Mean Photon-Number Constraint}\label{subsec:superadd_input}

The number of information bits that can be reliably communicated {\it per received photon at the channel output}---the photon information efficiency (PIE)---under a mean photon-number constraint per mode, $\E$, is given by $C(\E)/\E$ (nats/photon). From (\ref{eqn:capacity_energy1}), it can be shown that in order to achieve high PIE, $\E$ must be small. In the $\E\to 0$ regime, the capacity (\ref{eqn:capacity_energy1}) can be approximated as
\begin{equation}\label{eqn:capacity_scaling1}
C(\E)=\E\log\frac{1}{\E}+\E+o(\E),
\end{equation}
 which shows that  $C(\E)/\E\sim -\log \E$ for $\E \ll 1$. Thus there is no upper limit in principle to the photon information efficiency.

We will now show that in the high-PIE (low-photon-number) regime, the ultimate capacity~\eqref{eqn:capacity_scaling1} of optical channels under the mean photon-number constraint $\E$ is achievable closely even with a simple Binary Phase Shift Keying (BPSK) coherent-state constellation $\{\ket{\sqrt{\pmb\E}},\ket{-\sqrt{\pmb\E}}\}$, which satisfies the energy constraint with any prior distribution. The inner product between the two coherent states $\{\ket{\pmb\alpha},\ket{\pmb\beta}\}$ is, $\gamma = |\braket{\pmb\alpha}{\pmb\beta}|=\exp[{-|\pmb\alpha-\pmb\beta|^2/2}]$. Therefore,
$
|\braket{\sqrt{\pmb\E}}{-\sqrt{\pmb\E}}|=\exp[-2\E].
$
By plugging $\gamma=\exp[-2\E]$ into (\ref{eqn:cap_bin1_1}), we obtain the capacity of the BPSK-input constellation, which is denoted as $C_{\sf BPSK}(\E)$,
\begin{equation}
\begin{split}\label{CBPSK}
C_{\sf BPSK}(\E)
&=\E\log\frac{1}{\E}+\E+o(\E),
\end{split}
\end{equation}
which is equal to $C(\E)$ for the first- and second-order terms in the limit $\E\to 0$.
Therefore, in the low-photon-number limit, when $N\to\infty$, a binary constellation is enough to achieve $C(\E)$ up to the first two dominant terms of the capacity expansion.

For BPSK-output quantum states, the maximum achievable rate at $N=1$, which is denoted as $C_{1,\sf{BPSK}}(\E)$, can be calculated by using (\ref{eqn:C1_bin1}) as
\begin{equation}
\begin{split}\label{eqn:C1BPSK}
C_{1,\sf{BPSK}}(\E)&=2\E +o(\E).
\end{split}
\end{equation}
Thus, PIE of the BPSK channel caps off at $2$ nats/photon for $N=1$, while for $N$ large, achievable PIE $\to \infty$ as $\E\to 0$.
It would therefore be interesting to ask how large a JDR length $N$ is needed to bridge the gap between (\ref{eqn:C1BPSK}) and (\ref{CBPSK}) in the BPSK cq channel.

We now ask, for {\em arbitrary} binary coherent states under the same constraint on the mean photon number $\E$ per mode, how high an information rate is achievable when each mode is detected one-by-one, i.e., $N=1$.
The maximum capacity of binary coherent-state channel at $N=1$ under the mean photon-number constraint of $\E$ will be denoted as $C_{1,\sf{Binary}}(\E)$. 
This value $C_{1,\sf{Binary}}(\E)$ can be calculated in the regime $\E \to 0$ by finding the optimal binary states $\{\ket{\pmb\alpha_0},\ket{\pmb\alpha_1}\}$ and the input distribution $\{1-q, q\}$ that satisfies the mean photon-number constraint,
\begin{equation}
(1-q)|\pmb\alpha_0|^2+q|\pmb\alpha_1|^2\leq \E.
\end{equation}
The following lemma summarizes the result.

\begin{lem}\label{lem:binary}
{\it
The optimal binary inputs for $N=1$, are $\pmb\alpha_0=\sqrt{\E \cdot q^*/(1-q^*)}$ and $\pmb\alpha_1=-\sqrt{\E \cdot(1-q^*)/{q^{*}}}$ with 
\begin{equation}
q^*=\frac{\E}{2}\log\frac{1}{\E},
\end{equation}
and the resulting $C_{1,\sf{Binary}}(\E)$ is}
\begin{equation}\label{eqn:C1B}
C_{1,\sf{Binary}}(\E)=\E\log\frac{1}{\E}-\E\log\log\frac{1}{\E}+O(\E).
\end{equation}
\end{lem}
\begin{IEEEproof}
Appendix \ref{appen:lem1}.
\end{IEEEproof}

We conjecture that  for $C_1(\E)$,  a binary constellation is optimal in the low-photon-number limit. In other words, using an $M$-ary constellation, and a single-symbol receiver, one cannot beat the $C_{1, \sf Binary}(\E)$. We state it as the following conjecture.
\begin{con}{\it
When restricted to single-symbol measurements, the maximum achievable information rate under the mean photon-number constraint of $\E$ is the same as $C_{1,\sf Binary}(\E)$ up to the first two dominant terms, i.e., }
\beq
C_{1}(\E)=\E\log\frac{1}{\E}-\E\log\log\frac{1}{\E}+O(\E).
\eeq
\end{con}

Compared to the ultimate capacity (\ref{eqn:capacity_scaling1}), the first-order term of $C_{1,\sf{Binary}}(\E)$ in~\eqref{eqn:C1B} is the same as that of $C(\E)$. But, the difference in the second-order term shows how much less capacity is achievable at $N=1$ even with the optimized binary-output quantum states. In \cite{chung2011capacity}, we showed that (\ref{eqn:C1B}) can be achieved even using an on-off keying modulation $\left\{|{\pmb{0}\rangle}, |{\pmb{\alpha}\rangle}\right\}$ and a simple on-off direct-detection (photon counting) receiver. 
Therefore, in the context of optical communication in the high-PIE regime, the difference between the second-order terms of Eqs.~(\ref{eqn:capacity_scaling1}) and (\ref{eqn:C1B}) captures all of the performance gain from the complex quantum processing in the JDR.
Even though $C_{1,\sf{Binary}}(\E)$ and $C(\E)$ have the same leading term, in practice,  the two performances have significant difference. For example, if one wishes to achieve a photon efficiency of 10 bits/photon, one can solve for $\E_{\sf Holevo}$ that satisfies $C(\E)/\E= 10 $, and for $\E_{\sf 1,\sf{Binary}} $ that satisfies  $C_{1,\sf{Binary}} (\E)/\E= 10 $ bits/photon, respectively, and get $\E_{\sf Holevo} \approx 0.0027$ and $\E_{\sf 1,\sf{Binary}} \approx 0.00010$. This means that for $N=1$, after sending the average of $ 0.00010$ photons, a new input symbol should be modulated to achieve a photon efficiency of 10 bits/photon; whereas, for $N=\infty$ (Holevo limit) it is enough to transmit a new input symbol for every 0.0027 photons to achieve the same photon efficiency.
  Therefore, the  resulting spectral efficiencies differ by more than 1 order of magnitude. This example says that although (\ref{eqn:capacity_scaling1}) and (\ref{eqn:C1B}) have the same limit as $\E\to 0$, the rates at which this limit is approached are quite different, which is of practical importance. 
 As a result, the second term in the capacity results cannot be ignored. 
Therefore, it is  important to know how large a JDR length $N$ can bridge the gap in the second-order terms. We provide an answer for such questions in the following section.

\section{A Lower Bound on $C_N$}\label{sec4}
In this section, a lower bound is derived for the maximum achievable information rate at a finite blocklength $N$ of quantum measurements.
Using this bound, it is possible to calculate a blocklength $N$ at which a given fraction $0<\alpha\leq 1$ of the Holevo capacity of a pure-state cq channel can be achieved. Therefore, this result provides a framework to understand the trade-off between the (rate) performance and the (quantum) receiver complexity, for reliable transmission of classical information over a quantum channel.

\begin{thm}\label{thm:main1}
{\it For a pure-state classical input-quantum output (cq) channel $W: x \to \ket{\psi_x}$, $x \in {\mathcal X}$,
 the maximum achievable information rate using quantum measurements of blocklength $N$, which is $C_N/N$ as defined in (\ref{eqn:CNdef_sec1}), is bounded below as 
\begin{equation}\label{eqn:main1}
\frac{C_N}{N}\geq \max_R\left( \left(1-2e^{-N E(R)}\right)R-\frac{\log 2}{N}\right),
\end{equation}
where }
\begin{equation}\label{eqn:main2}
E(R)=\max_{0\leq s \leq 1}\left( \max_{P_X}\left(-\log\Tr(\rho^{1+s})\right)-sR\right),
\end{equation}
{\it with $\rho=\sum_{x \in {\mathcal X}} P_X(x) \ket{\psi_x}\bra{\psi_x}$.}
\end{thm}

By using this theorem, for the previously introduced BPSK  $\{\ket{\sqrt{\pmb\E}},\ket{-\sqrt{\pmb\E}}\}$ cq channel, a blocklength $N$ can be calculated at which the lower bound of (\ref{eqn:main1}) exceeds certain targeted rates below the capacity. In the previous section, it was shown that there is a great gap between $C_{1,\sf{BPSK}}(\E)/\E$ in (\ref{eqn:C1BPSK}) and $C_{\sf{BPSK}}(\E)/\E$ in (\ref{CBPSK}) as $\E \to 0$:
\begin{equation}
\begin{split}\nonumber
\frac{C_{1,\sf{BPSK}}(\E)}{\E}&=2+o(1),\\
\frac{C_{\sf{BPSK}}(\E)}{\E}&=\log\frac{1}{\E}+1+o(1).
\end{split}
\end{equation}
We saw that the capacity of the BPSK alphabet is as good as that of the optimal continuous Gaussian-distributed input as $N$ goes to infinity, i.e., $C_{\sf{BPSK}}(\E)$ is the same as $C(\E)$ in the first two dominant terms. However, at the measurement blocklength $N=1$, a BPSK constellation cannot even achieve the maximum mutual information of the optimal binary cq channel, $C_{1,\sf{Binary}}(\E)$ in (\ref{eqn:C1B}), and the PIE caps off at $2$ nats/photon. This means that the BPSK is far from the optimal constellation for $N=1$. Therefore, the performance of the BPSK channel depends significantly on the regime of $N$.
We will now find how much quantum processing is sufficient in order to communicate using the BPSK alphabet at rates close to its capacity.

Note that for the BPSK quantum states $\{\ket{\sqrt{\pmb\E}},\ket{-\sqrt{\pmb\E}}\}$, any input distribution satisfies the mean photon-number constraint of $\E$. Consequently, we can apply Theorem~\ref{thm:main1} to the BPSK channel---while automatically satisfying the mean photon-number constraint---even though the theorem itself does not assume any energy constraint. Let $C_{N,\sf{BPSK}}(\E)$ be the maximum achievable rate of BPSK cq channel under the mean photon-number constraint of $\E$ when the received quantum states are jointly measured by length-$N$ quantum measurements.
\begin{thm} \label{cor1}{\it For the coherent-state BPSK channel with the mean photon-number constraint of $\E\leq 0.01$, when the length of joint measurement $N \geq \E^{-1}\left(\log(1/\E)\right)$, we obtain
\begin{equation}\label{eqn:CNLBR}
\frac{C_{N,\sf{BPSK}}(\E)}{N}\geq \left( \left(1-2e^{-N \widetilde{E}(R^*)}\right)R^*-\frac{\log 2}{N}\right),
\end{equation}
where
\begin{equation}
\begin{split}\nonumber
&R^*=\E\log\frac{1}{\E}\left(1-\sqrt{\frac{\log\left(N\E\log(N\E)\right)}{N\E}}\right)+\E,\\
&\widetilde{E}(R)=-\log\left(\left(\frac{1+e^{-2\E}}{2}\right)^{1+s^*}+\left(\frac{1-e^{-2\E}}{2}\right)^{1+s^*}\right),\\
&\quad\quad\quad\;\;-s^*R,\\
&{\text{with}}\;\; s^*=\frac{\log\log(1/\E)-\log(R-\E)}{\log(1/\E)}-1.
\end{split}
\end{equation}
} 
\end{thm}
\begin{IEEEproof}
Appendix \ref{app:cor1}.
\end{IEEEproof}

Using this theorem, the following corollary can also be shown.
\begin{cor}\label{cornew1}{\it
For the coherent-state BPSK channel with the mean photon-number constraint of $\E\to 0$, at the measurement length of }
\begin{equation}\label{eqn:N11}
N=2\E^{-1}\left(\log(1/\E)\right)^{2}\left(\log\log(1/\E)\right)^{-1},
\end{equation}
{\it we can obtain}
\begin{equation}\label{eqn:CNBPSKB}
\frac{C_{N,{\sf BPSK}}(\E)}{N} \geq \E\log\frac{1}{\E}-\E\log\log\frac{1}{\E}+o\left(\E\log\log\frac{1}{\E}\right).
\end{equation}
{\it Moreover, at }
\begin{equation}\label{eqn:N12}
N=\E^{-1}\left(\log(1/\E)\right)^{2}\left(\log\log(1/\E)\right)^2,
\end{equation} 
{\it we can obtain the ultimate limit of PIE up to the first two dominant terms, }
\begin{equation}\label{eqn:CNBPSKBPSK}
\frac{C_{N,{\sf BPSK}}(\E)}{N}\geq \E\log\frac{1}{\E}+\E+o(\E).
\end{equation}
\end{cor}
\begin{IEEEproof}
Appendix \ref{app:cornew1}.
\end{IEEEproof}
\medskip
\begin{remark}\label{rem:N}
Note that this corollary shows that at the value of $N$ specified in (\ref{eqn:N11}), the coherent-state BPSK channel can attain the PIE at least as high as $C_{1,\sf{Binary}}(\E)/\E$  for the first- and second-order terms, which is the maximum achievable PIE at $N=1$ with the optimal binary-input satisfying the mean photon-number constraint $\E$.
Furthermore, at $N$ of (\ref{eqn:N12}), the lower bound already approaches $C_{\sf{BPSK}}(\E)$ and $C(\E)$ (to the first two order terms), which are the maximum information rates achievable with an arbitrarily large length of quantum processing for BPSK channel and for optimal continuous-input channel, respectively. It means that  in order to achieve the ultimate limit of PIE, we do not need to incorporate any further complicated quantum processing of which the measurement length is larger than $N$ in (\ref{eqn:N12}). 
\end{remark}
\medskip

\begin{figure}[t]
\centerline{\includegraphics[width=\columnwidth]{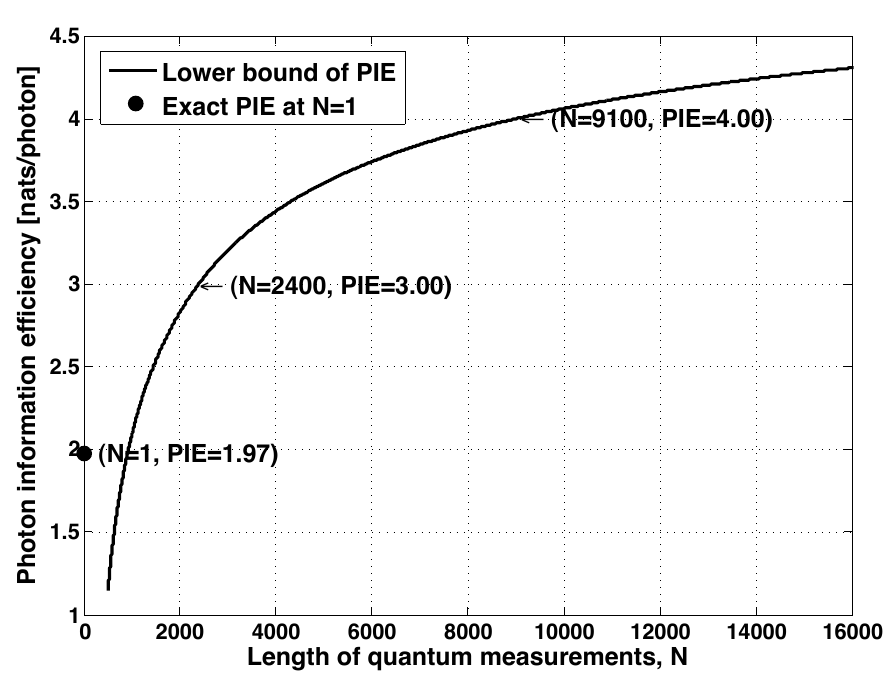}}
\caption{A lower bound on photon information efficiency of the BPSK channel $C_{N,\sf BPSK}(\E)/(N\E)$ at $\E=0.01 $ for the finite blocklength $N$.}
\label{fig:graph}
\end{figure}

Let us apply these results for the case when the average photon number transmitted per symbol, $\E$, is $0.01$. 
For $\E=0.01$, the inner product $\gamma:=|\braket{\sqrt{\pmb\E}}{-\sqrt{\pmb\E}}|=\exp[{-2\E}]=e^{-0.02}$. By plugging $\gamma$ into (\ref{eqn:cap_bin1_1}) and (\ref{eqn:C1_bin1}), and dividing the resulting capacities by $\E$, the PIE at an arbitrarily large $N$ is $5.55$ nats/photon, and at $N=1$, is $1.97$ nats/photon. Therefore, as $N$ increases from $1$ to $\infty$,  the gain in PIE from a joint measurement of an arbitrarily large length can be maximally $3.58$ nats/photon.

The photon information efficiency achievable by the BPSK channel at a finite measurement blocklength $N$, the right hand side of \eqref{eqn:CNLBR}, is plotted as a function of $N$ at $\E=0.01$ in Fig. \ref{fig:graph}.
From the lower bound on PIE in Fig. \ref{fig:graph}, it can be seen that at $N=2400$, a PIE of $3.0$ nats/photon can be achieved, and at $N=9100$, $4.0$ nats/photon is achievable. The lower bound is not tight in the regime of very small $N$, but it gets tighter as $N$ increases, and approaches the ultimate limit of PIE as $N\to\infty$.

Let us compare these  results with the approximations  from the scaling laws as $\E\to0$.
From the approximations in (\ref{CBPSK}), (\ref{eqn:C1BPSK}) and (\ref{eqn:C1B}), a few reference points of PIE are calculated at $\E=0.01$:
\begin{equation}
\begin{split}
&C_{\sf{BPSK}}(\E)/\E\approx\log(1/\E)+1=5.61 \; \text{nats/photon},\\
&C_{1,\sf{BPSK}}(\E)/\E\approx 2.00,\\
&C_{1,\sf{Binary}}(\E)/\E\approx\log(1/\E)-\log\log(1/\E)=3.08.
\end{split}
\end{equation}
We can see that these approximations of PIE for the BPSK channel at $N=1$ and $N=\infty$ are very close to the exact calculations. 
Moreover, it shows that when the optimal binary input for $N=1$ is used rather than the BPSK, the PIE of about $3.1$ nats/photon is achievable even at $N=1$.
The estimated length of $N$ to make the lower bound on $C_{N,{\sf BPSK}}(\E)$ be equal to the first two order terms of $C_{1,{\sf Binary}}(\E)$ is $N=2777$ from (\ref{eqn:N11}),  and $N$ to make the lower bound equal to the first two order terms of  $C_{\sf{BPSK}}(\E)$ is $N=4946$ from  (\ref{eqn:N12}). In Fig. \ref{fig:graph}, we showed that at $N=2400$, a PIE of $3.00$ nats/photon, which is close to $C_{1,\sf{Binary}}(\E)/\E$, is achievable. Therefore, the estimate of $N$ from (\ref{eqn:N11}) is quite accurate at $\E=0.01$. However, at $N=4946$, the achievable PIE from the lower bound of PIE is still $3.67$ nats/photon, which is $1.88$ nats/photon away from the maximum achievable PIE with an arbitrarily large length of quantum processing. Therefore, the estimate of $N$ in (\ref{eqn:N12}) is not very tight for $\E$ on the order of $10^{-2}$. It gets tighter for smaller $\E$ since the scaling laws are calculated in the limit of $\E\to 0$.

\section{Proof of Theorem~\ref{thm:main1}}\label{sec5}
Theorem \ref{thm:main1} will be proved based on two lemmas that will be introduced in this section. 
Note that in the definition of $C_N$ in (\ref{eqn:CNdef_sec1}), both the choice of the $N$-symbol inner encoder-JDR measurement pair, from which the superchannel distribution $p_{k|j}^{(N)}$ is determined, as well as the probability distribution over the inputs of the superchannel must be optimized, in order to find the maximum mutual information of the superchannel. 
The complexity of this optimization increases exponentially in $N$, and hence this optimization problem is intractable. Instead of trying to calculate the exact $C_N$, in Section~\ref{sec3}, we provided a lower bound on $C_N$ in the finite regime of $N$, which can be written as a simple optimization over single-letter input distribution. Therefore, we can easily calculate the lower bound on $C_N/N$ for any finite $N$. 

The proof of Theorem \ref{thm:main1} is based on two ideas: First, instead of tracking the exact superchannel distribution $p^{(N)}_{k|j}$, which depends on the detailed structure of the length-$N$ inner codewords and joint measurement, we focus on one representative quantity derived from $p^{(N)}_{k|j}$ that can be easily analyzed and optimized. Second, among superchannels that have the same value of this representative quantity, we find a superchannel whose mutual information is the smallest. The representative quantity is the average decoding error probability of the inner code  with a uniform distribution over the inner codewords and under the assumption that the cardinality of outputs of a quantum joint-detection receiver is equal to that of the inner-code messages. The average decoding error probability of the inner code is thus
\begin{equation}\label{eqn:peuni}
p_e=e^{-NR}\sum_{j=1}^{e^{NR}}\sum_{k\neq j}p_{k|j}^{(N)},
\end{equation}
where $R$ is the rate of the inner code.
We first summarize previous works that investigated  an upper  bound and a lower bound on $p_e$ over $N$-symbol inner encoder and JDR measurement pairs. We then provide a lower bound on the maximum mutual information of superchannel with a fixed $p_e$. Theorem \ref{thm:main1} will be proved by combining these two results.

\subsection{Achievability and Converse Results on the Average Probability of Error}
In ref.~\cite{holevo98arxiv}, Holevo showed the existence of a length-$N$ and rate-$R$ code that can achieve an $p_e$ exponentially decreasing in $N$.
\begin{lem}{}\label{lem:Holevo}
{\it
For a pure-state classical input-quantum output (cq) channel $W: x \to \ket{\psi_x}$, $x \in {\mathcal X}$, there exists a block code of length $N$ and rate $R$ that can be decoded by a set of measurements with the average probability of error satisfying
\begin{equation}\label{eqn:lwbdpe10}
p_e\leq 2\exp[-N E(R)],
\end{equation}
where, for $\rho=\sum_{x} P_X(x)\ket{\psi_x}\bra{\psi_x}$},
\begin{equation}\label{eqn:ER111}
E(R)=\max_{0\leq s\leq 1} \left[\max_{P_X}\left(-\log \Tr \left(\rho^{1+s}\right)\right)-sR\right].
\end{equation}
\end{lem}
Note that this result holds for $any$ positive integer $N$. Moreover, the exponentially decreasing rate of this upper bound on $p_e$ is characterized by the exponent $E(R)$ that is independent of $N$ and that can be calculated from the optimization over the single-letter input distribution $P_X$. 

Let us discuss the tightness of this upper bound on $p_e$ in terms of the exponentially decreasing rate of the bound as $N\to \infty$.
For a classical discrete memoryless channel (DMC), a lower bound on the average decoding error probability of block coding was first derived by Shannon-Gallager-Berlekamp in \cite{shannon1967lower}, and the bound is termed {\it sphere packing bound}. The sphere packing bound decreases exponentially with the blocklength $N$, and the exponent is tight at high rates below the capacity of the channel.

For quantum channels, an explicit lower bound on $p_e$ had not been established until very recently. In \cite{dalai2012sphere}, a quantum analogue of the sphere packing bound was first provided based on the idea of Nussbaum-Szkola mapping, introduced in \cite{nussbaum2009chernoff} as a tool to prove the converse part of the quantum Chernoff bound for binary hypothesis testing between two quantum states. The main result of \cite{dalai2012sphere} is summarized below. 

\begin{lem}[Sphere packing bound for quantum channels]
{{\it{ When we transmit classical information over a pure-state classical input-quantum output (cq) channel $W: x \to \ket{\psi_x}$, $x \in {\mathcal X}$, 
for every length-$N$ and rate-$R$ code, the average probability of error
\begin{equation}\label{eqn:lwbmax}
p_e\geq \exp[-N(E_{\sf{sp}}(R-\epsilon)+o(1))]
\end{equation}
for every $\epsilon>0$, where, for $\rho=\sum_x P_X(x)\ket{\psi_x}\bra{\psi_x}$, }}
\begin{equation}\label{eqn:EspR}
E_{\sf{sp}}(R)=\sup_{s\geq 0}\left(\max_{P_X}\left(-\log \Tr\left(\rho^{1+s}\right)\right)-sR\right).
\end{equation}
}
\end{lem}

From the lower bound in \eqref{eqn:lwbdpe10} and the upper bound in (\ref{eqn:lwbmax}), we can see that when $E(R)=E_{sp}(R)$, the exponent gets tight. It can be checked that $E(R)=E_{\sf{sp}}(R)$  in $R_0\leq R\leq C$ where $R_0$ is the rate at which the optimal $s$ achieving $E_{\sf{sp}}(R_0)$ in (\ref{eqn:EspR}) is equal to $1$.
Therefore, at high rates of $R$ where $R_0\leq R\leq C$, 
\begin{equation}\label{eqn:equER}
\limsup_{N\to \infty}-\frac{1}{N}\log p_e=E(R).
\end{equation}

\subsection{Equierror Superchannel}
Now, among superchannels $p_{k|j}^{(N)}$ that have the same value of $p_e$ defined in (\ref{eqn:peuni}), we find a superchannel whose mutual information is the smallest.
An {\it{equierror superchannel}}, which was first introduced in \cite{forney1966concatenated}, is defined with the following distribution:
\begin{equation}\label{eqn:superchannelsym}
\begin{split}
\overline{p}_{k|j}^{(N)}:= \left\{ 
  \begin{array}{l l}
    1-p_e, & \quad k=j\; ;\\
    \left(e^{NR}-1\right)^{-1}p_e, & \quad k\neq j.
  \end{array} \right.
\end{split}
\end{equation}
Note that this equierror superchannel satisfies (\ref{eqn:peuni}).
This channel assumes that the probability of making a decoding error for the inner code is equal for every input $j$, and when an error occurs, all wrong estimates $k\neq j$ can be observed with equal probabilities. Therefore, this channel is symmetric between inputs, and is symmetric between outputs except for the right estimate, i.e., $k=j$. 
Due to the symmetry, the input distribution that maximizes the mutual information of this channel is uniform. The resulting maximum mutual information of this equierror superchannel,
\begin{equation}
\begin{split}
\max_{p_j}I(p_j,\overline{p}_{k|j}^{(N)})&=NR-p_e\log\left(e^{NR}-1\right)-H_{\sf B} (p_e)\\
&> (1-p_e)NR-\log2,
\end{split}
\end{equation}
where $H_{\sf B}(p)=-p\log p -(1-p)\log(1-p)$. 

We will now show that the mutual information of this equierror superchannel is smaller than that of any other superchannels with the same value of the average decoding error probability, $p_e$.
\begin{lem}\label{lem:equierror}
{\it 
For any $p_{k|j}^{(N)}$ with a fixed $p_e$ defined in (\ref{eqn:peuni}),
\begin{equation}
\max_{p_j}I\left(p_j,p_{k|j}^{(N)}\right) \geq \max_{p_j}I\left(p_j,\overline{p}_{k|j}^{(N)}\right)
\end{equation}
for the equierror superchannel, $\overline{p}_{k|j}^{(N)}$ with the same $p_e$.
}
\end{lem}
\begin{IEEEproof}
For a random variable $X$ that is uniformly distributed over $e^{NR}$ inputs, and the conditional distribution $P_{Y|X}(k|j):=p^{(N)}_{k|j}$,
\begin{equation}
\begin{split}\label{eqn:pf1}
\max_{p_j}I\left(p_j,p_{k|j}^{(N)}\right)&\geq I(X;Y)=NR-H(X|Y).
\end{split}
\end{equation}
From the Fano's inequality, we have
\begin{equation}
\begin{split}\nonumber
H(X|Y)&\leq H_B\left(\Pr(X\neq Y)\right) +\Pr(X\neq Y) \log(e^{NR}-1)\\
&=H_B(p_e)+p_e\log\left(e^{NR}-1\right).
\end{split}
\end{equation}
By combining the above two inequalities, we get
\begin{equation}
\begin{split}
\max_{p_j}I\left(p_j,p_{k|j}^{(N)}\right)&\geq  NR-p_e\log\left(e^{NR}-1\right)-H_B(p_e)\\
&=\max_{p_j}I\left(p_j,\overline{p}_{k|j}^{(N)}\right).
\end{split}
\end{equation}
\end{IEEEproof}

Then, by the definition of $C_N$  in (\ref{eqn:CNdef_sec1}) and Lemma \ref{lem:equierror}, when there exists an inner code of length $N$ and rate $R$ that can be decoded by a set of length-$N$ measurements with an average decoding error probability $p_e$, it can be shown that
\begin{equation}\label{eqn:thmpfpart1}
\frac{C_N}{N}\geq \max_{p_j}\frac{I\left(p_j,p_{k|j}^{(N)}\right)}{N}>(1-p_e)R-\frac{\log 2}{N}.
\end{equation} 
By combining Lemma~\ref{lem:Holevo} with (\ref{eqn:thmpfpart1}), we get
\begin{equation}
\frac{C_N}{N}\geq \left( \left(1-2e^{-N E(R)}\right)R-\frac{\log 2}{N}\right),
\end{equation}
with $E(R)$ in (\ref{eqn:ER111}) for any $R>0$.
By maximizing the right hand side over the rate $R$, we get Theorem~\ref{thm:main1}.

\section{A Unifying Framework for $C_N$ of a Quantum Channel and $C_N^{(c)}$ of a Classical Channel}\label{sec6}

In Section~\ref{sec3}, we demonstrated strict superadditivity of $C_N$ by showing that $C_1<C$ for binary pure-state quantum channels with and without an energy constraint. We provided a general lower bound on $C_N/N$ for a fixed $N$ in Theorem~\ref{thm:main1}, which made it possible for us to understand the trade-off between the maximum achievable information rate and the complexity of quantum processing at the receiver as $N$, the length of the joint measurement, increases.

The superadditivity of $C_N$ of a cq channel is traditionally interpreted in the following way. A set of length-$N$ joint-detection quantum measurements can induce a classical superchannel whose transition probability matrix does not factor into a product conditional distribution over the $N$ uses of the quantum channel, despite the fact that the inputs to the underlying cq channel, and the action of the cq channel itself are independent over each channel use. The Shannon capacity of this induced classical superchannel can be higher than $N$-times the Shannon capacity of the classical channel induced by pairing the memoryless cq channel with the optimal symbol-by-symbol receiver measurement. This capability of inducing a classical superchannel by harnessing the optimally-correlated quantum noise in the $N$-fold Hilbert space of the product-state quantum codeword is what increases the number of information bits extractable per modulation symbol, when a longer block of symbols is detected collectively {\em while the modulated symbols of the $N$-length codeword are still in the quantum (optical) domain}. This is an example of what is known as {\em non-locality without entanglement} in quantum mechanics~\cite{bennett1999}, where ``non-local" (or, joint) measurements can perform better despite the fact that the systems being measured are in a product (non-entangled) quantum state.

Despite the fact that the above intuition of why superadditivity appears in the capacity of quantum channels is somewhat satisfying, this does not provide enough quantitative insight to analyze how $C_N$ increases with the length $N$ of a joint measurement. 
In this section, we will introduce a quantitative insight to interpret the lower bound on $C_N/N$ of a cq channel provided in Theorem~\ref{thm:main1} by establishing a similar type of lower bound on $C_N^{(c)}/N$ of a classical DMC, where $C_N^{(c)}/N$ is the asymptotic capacity of a classical DMC where the inner decoder is restricted to making hard estimates at a finite blocklength $N$ for the inner-code messages of cardinality $M\leq \lfloor e^{NC^{(c)}}\rfloor$.
This latter setting requires a generalization of previous work by Forney~\cite{forney1966concatenated} to the case of a fixed inner-code blocklength $N$.

We will see in this section that a quantum receiver having to produce a classical output by jointly detecting up to an $N$-length modulated codeword block at the output of a cq channel is mathematically analogous to a classical inner decoder having to make a {\em hard decision} on messages encoded into $N$-channel-use-long codewords. Our formulation lends a quantitative insight on both problems, which allows us to calculate a lower bound on $C_N/N$ as stated in Theorem~\ref{thm:main1}, as well an analogous lower bound for the analogous classical problem, which we discuss next.

\subsection{A Lower Bound on $C_N^{(c)}/N$}
In this section, we examine  the asymptotic capacity $C_N^{(c)}/N$ in~\eqref{eqn:classCNhard1} of a classical DMC with superposition coding at a finite inner-code blocklength $N$ where the inner decoder makes a hard decision for the inner-code message, which is selected from the set of cardinality $M\leq \lfloor e^{NC^{(c)}}\rfloor$.
More specifically, we provide a lower bound on $C_N^{(c)}/N$ for a finite $N$. This result will then be corroborated with the lower bound on $C_N/N$ of a cq channel under a unified framework.

In~\cite{forney1966concatenated}, Forney showed that even if there is some loss of information at the inner decoder by making a hard decision on the message of the inner code, as the inner-code blocklength $N$ as well as the outer-code blocklength $n$ go to infinity, the capacity of the underlying classical DMC can be achieved with concatenated coding. 
This result was proved by analyzing a lower bound on the maximum mutual information of superchannels where the outputs of the superchannel are the estimate of the inner-code message.
To get the lower bound, the equierror superchannel defined in (\ref{eqn:superchannelsym}), whose mutual information is smaller than that of any other superchannel with the same $p_e$, was used.
The average probability of decoding error $p_e^{(c)}$ for the inner-code message over a classical DMC can be analyzed by using the error exponent of the classical DMC $P_{Y|X}$ in \cite{gallager1968information}.
By using the random coding arguments, it can be shown that there exists an inner code of length $N$ and rate $R$ with the average decoding error probability $p_e^{(c)}$ as low as
\begin{equation}\label{eqn:optp}
p_e^{(c)}=\exp[-N(E^{(c)}(R)+o(1))]
\end{equation}
for $N\to \infty$, when 
\begin{equation}\label{eqn:ERclassic}
E^{(c)}(R)=\max_{0\leq s \leq 1}\left(\max_{P_X}\left(E^{(c)}_0(s,P_X)\right)-sR\right)
\end{equation}
with
\begin{equation}\label{eqn:E0}
E^{(c)}_0(s,P_X):=-\log \sum_{y}\left[\sum_{x} P_X(x) P_{Y|X}(y|x)^{\frac{1}{1+s}}\right]^{1+s}.
\end{equation}
Here the superscript $(c)$ stands for classical channels to avoid confusion with the previous result of quantum channels. 
By using the $p^{(c)}_e$ in (\ref{eqn:optp}) for analyzing the maximum mutual information of the equierror channel, it can be shown that the capacity of the DMC, which is $C^{(c)}=\max_{P_X}I(P_X,P_{Y|X})$, is achievable by the concatenated code {\it as both the inner-code blocklength $N$ and the outer-code blocklength $n$ go to infinity}, even when the inner decoder makes hard decisions on estimating the inner-code messages, and discards all the rest of the information about the channel output. The loss of information at the inner decoder, however, degrades the achievable error exponent over all rates below capacity.

We now establish a lower bound on $C_N^{(c)}/N$ in~\eqref{eqn:classCNhard1} for a fixed inner-code blocklength $N$.
By using Lemma~\ref{lem:equierror}, it can also be shown that when there exists a code of length $N$ and rate $R$ whose probability of decoding error is $p^{(c)}_e$,
\begin{equation}
\begin{split}\label{eqn:comb1}
\frac{C^{(c)}_N}{N}&>(1-p^{(c)}_e)R-\frac{\log2}{N}.
\end{split}
\end{equation}
Moreover, in \cite{gallager1968information}, it is shown that for the classical DMC $P_{Y|X}$, there exists a code of length $N$ and rate $R$ whose probability of error $p^{(c)}_e$ is bounded above as
\begin{equation}\label{eqn:comb2}
p^{(c)}_e\leq \exp[-NE^{(c)}(R)]
\end{equation}
with $E^{(c)}(R)$ in (\ref{eqn:ERclassic}). By combining (\ref{eqn:comb1}) and (\ref{eqn:comb2}), the following theorem can be demonstrated.
\begin{thm}\label{thm:57class} {\it With a fixed inner-code blocklength $N$, 
\begin{equation}\label{eqn:CNlbclassic}
\frac{C^{(c)}_N}{N}\geq \max_{R}\left(\left(1-e^{-NE^{(c)}(R)}\right)R-\frac{\log2}{N}\right),
\end{equation}
with $E^{(c)}(R)$ as defined in (\ref{eqn:ERclassic}).
}
\end{thm}

Note that the lower bound on $C^{(c)}_N/N$ in (\ref{eqn:CNlbclassic}) strictly increases with $N$ and approaches $C^{(c)}$ as $N\to\infty$. Moreover, it has exactly the same form as that of the quantum channel in (\ref{eqn:main1}) except for the difference in $E^{(c)}(R)$ and a constant multiplying $e^{-NE^{(c)}(R)}$. 
The reason why $C^{(c)}_N/N$ is strictly smaller than the Shannon capacity  $C^{(c)}$ for a finite inner-code blocklength $N$ is because the hard decision at the inner decoder results in a significant amount of loss of information, which hurts the communication rate. As $N$ increases, the quality of the hard decision is improved, which makes it possible to achieve a progressively higher information rate.

An analogous line of reasoning could also be applied to provide a lower bound on $C_N/N$ in the quantum channel as stated in Theorem~\ref{thm:main1}, by replacing the role of inner decoder with a quantum joint-detection receiver that necessarily makes a classical output on finite blocks of quantum states. Therefore, as opposed to the original explanation of superadditivity, which  regards it as a gain from joint measurement of quantum states, our formulation rather focuses on the information loss resulting from the quantum joint detection at a finite blocklength $N$ in order to derive a useful general lower bound on $C_N/N$ for a fixed $N$.

One thing that would be good to note here is that the quantum JDR acting on the $N$-length inner (quantum) codeword does not have to generate a hard-decision output on the inner-code message. In fact, it is known that the number of outcomes in the POVM that maximizes the accessible information for $M$ linearly independent pure states, grows as $O(M^2)$~\cite{Dav78}. In our case, $M = e^{NR}$. In recent years, some quantum decoding techniques have been developed---such as the sequential decoder~\cite{giovannetti2012achieving} and the quantum successive-cancellation decoder for a quantum polar code~\cite{Wil12a, Wil12b}---that achieve the Holevo capacity, which make weak (partially-destructive) measurements on the received codeword, and retain the post-measurement states for further conditional quantum processing. Recently Wilde {\em et al.} used a quantum version of the likelihood ratio test, originally proposed by Fuchs and Caves~\cite{Fuc95}---another non-destructive quantum measurement---in an attempt to build an efficient decoder for the quantum polar code~\cite{Wil13}. However, all these weak non-destructive quantum measurements are very hard to realize in practice.

\subsection{An Approximation of the Lower Bounds on $C_N/N$ and on $C_N^{(c)}/N$}

We will simplify the lower bounds on $C_N/N$ in~\eqref{eqn:main1} and on $C^{(c)}_N/N$ in~\eqref{eqn:CNlbclassic} by finding an approximation of the error exponent $E(R)$ in~\eqref{eqn:main2} for the quantum channel and $E^{(c)}(R)$ in~\eqref{eqn:ERclassic} for the classical DMC, respectively. Using the simplified lower bounds, it will be possible to compare the quantum channel and the classical channel by calculating the inner-code blocklength $N$ sufficient to achieve a given fraction of the ultimate capacity of each channel. 
To avoid confusion, from this point on, a function for the quantum channel will be written with a superscript $(q)$ and that for the classical DMC with a superscript $(c)$; for example, $E^{(q)}(R)$ and $E^{(c)}(R)$.

The error exponent of the classical DMC, $E^{(c)}(R)$ in (\ref{eqn:ERclassic}), can be approximated by the Taylor expansion at the rate $R$ close to the capacity $C^{(c)}$ as
\begin{equation}\label{eqn:appEr49}
E^{(c)}(R)=\frac{1}{2V^{(c)}}\left(R-C^{(c)}\right)^2+O\left(\left(R-C^{(c)}\right)^3\right),
\end{equation}
with a parameter $V^{(c)}$, where
\begin{equation}\label{eqn:Vc}
\begin{split}
&V^{(c)}=\\
&\sum_{x,y}p_x p_{y|x} \left[\left(\log\frac{p_{y|x}}{p_{y}}-\left(\sum_{x',y'} p_{x'} p_{y'|x'} \log\frac{p_{y'|x'}}{p_{y'}}\right)\right)^2 \right],
\end{split}
\end{equation}
for the capacity achieving input distribution $p_x:=P_X^*(x)$ and the corresponding capacity achieving output distribution $p_y:=P_Y^*(y)$ according to the channel $p_{y|x}:=P_{Y|X}(y|x)$.
In (\ref{eqn:Vc}), $V^{(c)}$ is the variance of $\log (p_{y|x}/p_y)$ under the distribution $p_x p_{y|x}$, and was termed the {\it channel dispersion} in \cite{polyanskiy2010channel}.

Similarly, the error exponent of the quantum channel, $E^{(q)}(R)$ in (\ref{eqn:main2}), can be approximated with a parameter $V^{(q)}$, which is a characteristic of the quantum channel similar to the channel dispersion of the classical DMC. The definition of $V^{(q)}$ depends on the average density operator $\rho$, which fully characterizes the classical capacity of the pure-state quantum channel.
For a set of quantum states $\{\ket{\psi_x}\}$, when $P_X^*$ is the optimal input distribution that attains the capacity of the quantum channel $C^{(q)}=\max_{P_X}\Tr(-\rho\log\rho)$ where $\rho=\sum_x P_X(x)\ket{\psi_x}\bra{\psi_x}$, the parameter $V^{(q)}$ is defined by the eigenvalues of the density operator $\rho$ at $P_X=P_X^*$.
Let us denote the eigenvalues of $\rho$ by $\sigma_i$, $i=1,\dots, D$, where $D$ is the dimension of the space spanned by the quantum states $\{\ket{\psi_x}\}$.
From the fact that $\rho$ is a positive operator and $\Tr(\rho)=1$, it can be shown that each $\sigma_i\geq 0$ for all $i$ and $\sum_{i=1}^D \sigma_i=1$.
Then, $V^{(q)}$ is defined as a variance of the random variable $(-\log\sigma)$ where $\sigma\in\{\sigma_i\}$ with probability distribution $\{\sigma_1,\dots,\sigma_D\}$, i.e.,
\begin{equation}\label{eqn:Vforq}
V^{(q)}=\sum_{i=1}^D \sigma_i (-\log \sigma_i)^2-\left(\sum_{i=1}^D \sigma_i \left(-\log\sigma_i\right)\right)^2.
\end{equation}
This quantity was first defined in~\cite{wilde2014second}.
By the Taylor expansion of $E^{(q)}(R)$ in (\ref{eqn:main2}) at the rate $R$ close to $C^{(q)}$, it can be shown that 
\begin{equation}\label{eqn:ERTaylor}
E^{(q)}(R)=\frac{1}{2V^{(q)}}\left(R-C^{(q)}\right)^2+O\left(\left(R-C^{(q)}\right)^3\right).
\end{equation}
Therefore, both the error exponent of the classical DMC and that of the quantum channel can be approximated as a quadratic term in the rate $R$ with the quadratic coefficient inversely proportional to the dispersion of the channel. Since the lower bounds on $C^{(q)}_N$ and on $C^{(c)}_N$  as well as the approximated error exponents of $E^{(q)}(R)$ and of $E^{(c)}(R)$ have similar forms, it is possible to compare the classical DMC and the quantum channel by a common simplified lower bound on $C^{(q)}_N$ and on $C^{(c)}_N$, which can be written with the parameters $\left(V^{(q)},C^{(q)}\right)$ and $\left(V^{(c)},C^{(c)}\right)$, respectively, as follows.
\begin{thm}\label{thm:58qc} {\it For both a classical DMC for which $(C_N,V,C)=\left(C^{(c)}_N,V^{(c)},C^{(c)}\right)$ and a pure-state classical-quantum channel  for which $(C_N,V,C)=\left(C^{(q)}_N,V^{(q)},C^{(q)}\right)$, when the channel dispersion $V$ and the capacity $C$ satisfy \rmnum 1) $\sqrt{\frac{V}{NC^2}}\to 0$ as $N\to \infty$ and \rmnum 2) $V\cdot C$ is finite, the maximum achievable information rate at the inner-code blocklength $N$ is bounded below as
\begin{equation}
\begin{split}\label{eqn:CNbdV}
\frac{C_N}{N}&\geq C\cdot \left(1-\sqrt{\frac{V}{NC^2}\log\left(\frac{NC^2}{V}\right)}\right)\\
&\quad-\frac{\log2}{N}+O\left(\sqrt{\frac{V}{N\log({NC^2}/{V})}}\log\log\left(\frac{NC^2}{V}\right)\right).
\end{split}
\end{equation}
}
\end{thm}
\begin{IEEEproof}
The quadratic approximation of $E(R)$ can be used to find a simplified form for a lower bound on $C_N/N$.
Both for the quantum channel and the classical channel, $C_N$ is lower bounded by
\begin{equation}\label{eqn:mainrep1}
\frac{C_N}{N}\geq \max_{R} \left(\left(1-2e^{-NE(R)}\right)R-\frac{\log2}{N}\right),
\end{equation}
from Theorems~\ref{thm:main1} and~\ref{thm:57class}.
Then, for a fixed rate
\begin{equation}\label{eqn:Rstar}
R^*=C\cdot \left(1-\sqrt{\frac{V}{NC^2}\log\left(\frac{NC^2}{V}\log\frac{NC^2}{V}\right)}\right),
\end{equation}
 the approximated error exponent at $R^*$ is
\begin{equation}
\begin{split}\label{eqn:Ocdot}
&E(R^*)=\frac{1}{2N}\log\left(\frac{NC^2}{V}\log\frac{NC^2}{V}\right) \\
&\qquad\quad+O\left(\frac{VC}{N}\sqrt{\frac{V}{NC^2}}\left(\log\left(\frac{NC^2}{V}\log\frac{NC^2}{V}\right) \right)^{3/2}\right)
\end{split}
\end{equation}
from~\eqref{eqn:appEr49} for a classical DMC and~\eqref{eqn:ERTaylor} for a quantum channel, respectively. 
It can be checked that under the assumptions of \rmnum 1) $\sqrt{\frac{V}{NC^2}}\to 0$ and \rmnum 2) $V\cdot C$ being finite, the term in $O(\cdot)$ in (\ref{eqn:Ocdot}) approaches 0 as $N\to\infty$, which results in
\begin{equation}
\begin{split}\label{eqn:eNERap}
e^{-NE(R^*)}&=\sqrt{\frac{V}{NC^2\log\left({NC^2}/{V}\right)}}\,\left(1+o(1)\right).
\end{split}
\end{equation}

By plugging (\ref{eqn:Rstar}) and (\ref{eqn:eNERap}) into the lower bound (\ref{eqn:mainrep1}), $C_N/N$ can be bounded below as shown in (\ref{eqn:CNbdV}).
\end{IEEEproof}

\medskip
\begin{remark}
{ From the lower bound of Theorem~\ref{thm:58qc}, we can see that the inner-code blocklength $N$ at which the lower bound is equal to a given constant fraction of the capacity is proportional to $V/C^2$.
In \cite{polyanskiy2010channel}, a different problem of analyzing the maximal channel coding rate for a classical DMC at a given blocklength $N$ and error probability $\epsilon$ was investigated.
With $M^{(c)}_{N,\epsilon}$ denoting the maximum number of messages that can be transmitted over a finite number ($N$) of channel uses with average error probability $\epsilon$, it was shown that
\beq
\frac{\log M^{(c)}_{N,\epsilon}}{N}=C\left(1-Q^{-1}(\epsilon)\sqrt{\frac{V}{NC^2}}\right)+O\left(\frac{\log N}{N}\right)
\eeq
where $Q^{-1}$ is the inverse of the $Q$-function, the tail probability of the standard normal distribution.
This result shows that to achieve a $\eta$-fraction of the capacity $C$ with error probability $\epsilon$, the required block length $N$ is again proportional to $V/C^2$. 
Therefore, even though $C_N^{(c)}/N$ (error-free bits per channel use) and $(\log M^{(c)}_{N,\epsilon})/{N}$ (channel coding rate with error probability $\epsilon$) consider different scenarios, the quantity $V/C^2$ appear for both of the problems as a parameter that governs the second-order asymptotics of the channel behavior. 

}
\end{remark}
\medskip

Since the same bound on $C_N/N$ as in (\ref{eqn:CNbdV}) holds both for the quantum and the classical channels, using the parameter $V/C^2$, we can compare the behavior of the quantum channel and of the classical DMC. 
For the BPSK quantum channel, by using the two eigenvalues of $\rho$ at $P_X^*$, which are $\sigma_1=(1-e^{-2\E})/2$ and $\sigma_2=(1+e^{-2\E})/2$, the channel dispersion in~\eqref{eqn:Vforq} and the capacity can be calculated as
\begin{equation}
\begin{split}
V_{\sf{BPSK}}^{(q)}&=\E\left(\log\frac{1}{\E}\right)^2(1+O(\E)), \,{\rm and}\\
C_{\sf{BPSK}}^{(q)}&=\E\log\frac{1}{\E}+\E+o(\E).
\end{split}
\end{equation}
Then, $V_{\sf{BPSK}}^{(q)}/{(C_{\sf{BPSK}}^{(q)})^2}\approx 1/\E$ for the low-photon-number regime where $\E \to 0$.
For the classical additive white Gaussian noise (AWGN) channel in the low-power regime where ${\sf SNR}\to 0$, $V_{\sf{AWGN}}^{(c)}/(C^{(c)}_{\sf{AWGN}})^2$ can be calculated by using the result of \cite{polyanskiy2010channel}, and it is $4/{\sf SNR}$. For both channels, $V/C^2$ is thus inversely proportional to the energy to transmit the information per channel use. This means that as the energy per channel use decreases, in order to make the lower bound meet a targeted fraction of capacity, it is necessary to adopt a longer inner code.

\section{Discussions}\label{sec:6con}

The Holevo capacity of a classical-quantum (cq) channel, i.e., the ultimate rate of reliable communication for sending classical data over a quantum channel using product-state codewords, is a doubly-asymptotic result; meaning the achievability of the capacity $C$ has been proven so far for the case when the transmitter is allowed to code over an arbitrarily large sequence of quantum states (spanning $N_c$ channel uses), {\em and} when the receiver is assumed to be able to {\it jointly} measure quantum states of the received codewords, also over $N_c$ channel uses, while $N_c \to \infty$. The assumption that arbitrarily large number of quantum states can be jointly measured (using a potentially very complicated quantum joint operation) is impractical for realizations of joint-detection receivers---particularly in the context of optical communication. Our goal in this paper was to separate these two infinities: the coding blocklength $N_c$ (a relatively inexpensive resource), and the length of the joint-detection receiver, $N \le N_c$ (a far more expensive resource), and to evaluate how the capacity $C_N$, constrained to length-$N$ joint measurements (but no restrictions on the classical code complexity), grows with $N$.
We analyzed superadditivity in capacity of a pure-state classical input-quantum output channel while focusing on the quantitative trade-off between reliable-rate performance and quantum-decoding complexity.
In order to analyze this trade-off, we adopted a concatenated coding scheme where a quantum joint-detection receiver acts on finite-blocklength quantum codewords (viz., a train of $N$ modulated laser-light pulses) of the inner code, and we found a lower bound on the maximum achievable information rate $C_N/N$ as a function of the length $N$ of the quantum measurement that decodes the inner code.

We also defined and studied the information rate $C^{(c)}_N/N$ achievable over a classical discrete memoryless channel (DMC), when a concatenated coding scheme is employed with the inner decoder forced to make hard decisions on $N$-length inner codewords (with no restriction on the block length of the outer code) when the cardinality of the inner-code message set is limited to $M\leq \lfloor e^{NC^{(c)}}\rfloor$. We showed that this information rate $C_N^{(c)}$ also exhibits superadditivity. The superadditivity in the case of the classical problem arises due to a loss of information from the hard decisions at the inner decoder made on finite blocklength inner codes as well as the limited cardinality of the input set. Even though the superadditivity in the quantum channel is not all due to the loss of information, this viewpoint could also be applied for the quantum channel in order to provide a general lower bound on $C_N/N$.
We developed a unifying framework, within which the superadditivity in capacity of the classical DMC and that of the pure-state classical input-quantum output channel can both be quantified by a parameter $V/C^2$ (where $V$ is the channel dispersion, and $C$ the channel capacity, of the respective problem), in the sense that it is proportional to the inner-code measurement length $N$ that is sufficient to achieve a given fraction of the respective asymptotic capacity.

In this paper, when we discussed the superadditivity of coherent-state channels with the photon-number constraint, we focused our discussion on the low-photon-number regime.
It is known that coherent-state (ideal laser light) modulation is sufficient to achieve the Holevo capacity at any power (photon number) regime, using a coherent-state random code constructed by choosing the coherent-state amplitude of each symbol of each codeword from a circularly-symmetric complex Gaussian distribution. The above was proven for pure-loss channels in Ref.~\cite{giovannetti2004classical}, and was recently extended to the case of the lossy-noisy bosonic channel in Ref.~\cite{patron2014}. For the pure-loss channel, the gap between the Holevo capacity and the Shannon capacities associated with various conventional optical receivers (viz., homodyne, heterodyne, and direct-detection) widens in the low-photon-number regime (see Fig. 6 of ~\cite{Tak14}). In the high-photon-number regime on the other hand, heterodyne detection is known to be asymptotically capacity optimal. In other words, $\lim_{{\cal E} \to \infty}(C({\cal E})/\log(1+{\cal E})) = 1$, where $C({\cal E})$ is the Holevo capacity in bits per mode for the pure-loss channel with ${\cal E}$ photons received per mode, and the Shannon capacity achieved by an ideal (local-oscillator shot-noise-limited) Heterodyne detection is $\log(1+{\cal E})$ bits per mode. Therefore, given heterodyne detection makes symbol-by-symbol measurements, there is not much gap between $C_1(\E)$ and $C(\E)$ in the high-photon-number regime, and hence there is not much room for superadditivity in capacity. However, in the high-photon-number regime, non-standard, but single-symbol measurements may outperform standard optical receivers in the error exponent in discriminating symbols of a modulation constellation (in other words achieve a significantly lower channel dispersion), which would translate to a superior finite blocklength rate achievable by these non-standard receivers, even though heterodyne detection is capacity-optimal in this regime~\cite{Tan15}.

Finally, we hope that the problem setup proposed in this paper, which motives the study of classical communication over quantum channels with a finite-length joint measurement, will spur further developments to find tighter bounds on $C_N/N$, which might reveal more quantitative insights to fully understand the superadditivity phenomenon. 
In this paper,  in order to provide a lower bound on $C_N/N$ we focused on the information loss caused by the finite-length joint measurement that generates hard estimates on the inner-code message.
This perspective allowed us to develop a unifying framework between a quantum channel and a classical DMC when we analyzed lower bounds on $C_N/N$ of a quantum channel and on $C_N^{(c)}/N$ of a classical DMC. 
Even though this connection provided us a useful lower bound on $C_N/N$, this lower bound does not capture unique properties of quantum channels that never appear in classical DMCs. 
Therefore, as one of the reviewers of this paper has suggested, it would be very interesting if future studies can find new bounds on $C_N /N$, which would shed some additional light on how fast this quantity approaches the Holevo limit $C$ as $N$ increases in terms of some measure of ``non-classicality'' of the quantum channel. 
In this paper, we studied the superadditivity in capacity of a pure-state cq channel for a particular example of such a channel, which maps a complex input number to a coherent state.
This channel is relevant in practice for quantum optical communications.
However, it would be more interesting if one can extend the study of $C_N/N$ for a general classical input-quantum output channel. 
Lastly, we hope that our problem setup will not only motivate further analysis of $C_N/N$ with $N$-mode joint measurements, but also lead us to better understand the mathematical structure of such $N$-length joint measurements to turn them into recipes for structured designs for optical joint-detection receivers. 


\ifCLASSOPTIONcaptionsoff
  \newpage
\fi

\appendices

\section{Proof of Lemma \ref{lem:binary}}\label{appen:lem1}
We denote $C_{1,\sf{Binary}}(\E)$ as the maximum achievable information rate of a  binary cq channel paired with the optimal measurement of length $1$, under the mean photon-number constraint of $\E$. Binary quantum states $\{\ket{\pmb\alpha_0},\ket{\pmb\alpha_1}\}$ with input distribution $\{1-q,q\}$ should satisfy the mean photon-number constraint of $(1-q)|\pmb{\alpha_0}|^2+q|\pmb{\alpha_1}|^2\leq\E$. For this binary cq channel, the number of outcomes of the optimal POVM can be restricted to two without hurting the maximum mutual information~\cite{Dav78}, and as shown in~\cite{ban1999cut} the maximum mutual information of the resulting binary-input binary-output channel with the optimal length-1 POVM can be written as
\begin{equation}\label{eqn:C1binref}
\begin{split}
&C_{1,\sf{Binary}}(\E)=\\
&\max_{\{(q,\pmb\alpha_0,\pmb\alpha_1):(1-q)|\pmb{\alpha_0}|^2+q|\pmb{\alpha_1}|^2=\E\}} \left(H_{\sf B}(q) -H_{\sf B}(p(q,\pmb{\alpha_0},\pmb{\alpha_1}))\right)
\end{split}
\end{equation}
where $H_B(x)=-x\log x-(1-x)\log(1-x)$, and
\begin{equation}\label{eqn:p}
p(q,\pmb{\alpha_0},\pmb{\alpha_1})=\frac{1-\sqrt{1-4q(1-q)e^{-|\pmb\alpha_0-\pmb\alpha_1|^2}}}{2}.
\end{equation}
In \cite{sohma2000binary}, an approximation of $C_{1,\sf{Binary}}(\E)$ was calculated up to the first dominant term. Here we first summarize the result of \cite{sohma2000binary}, which shows the process of finding the optimal $\{\ket{\pmb\alpha_0},\ket{\pmb\alpha_1}\}$ for a fixed $q$. After that, we provide a more accurate approximation of $C_{1,\sf{Binary}}(\E)$ than that of \cite{sohma2000binary} up to the first two dominant terms by finding the optimal $q$.

Let us first find the optimal output states $\{\ket{\pmb\alpha_0},\ket{\pmb\alpha_1}\}$ for a fixed $q$. To minimize $H_{\sf B}(p(q,\pmb{\alpha_0},\pmb{\alpha_1}))$ for a fixed $q$, we need to maximize  $|\pmb\alpha_0-\pmb\alpha_1|^2$ under the energy constraint $(1-q)|\pmb{\alpha_0}|^2+q|\pmb{\alpha_1}|^2\leq\E$. To maximize $|\pmb\alpha_0-\pmb\alpha_1|^2$, $\pmb\alpha_0$ and $\pmb\alpha_1$ should have a relationship such that
\begin{equation}\label{eqn:a1a0k}
\pmb\alpha_1=-k\pmb\alpha_0 
\end{equation}
for a real number $k\geq 0$ satisfying
\begin{equation}\label{eqn:a1a0ke}
(1-q)|\pmb\alpha_0|^2+q|\pmb\alpha_1|^2=(1-q+k^2\cdot q)|\pmb\alpha_0|^2=\E.
\end{equation} 
The reason why we can restrict $k$ to be a real number is simple. For two coherent states, we can alway choose an axis passing through them and consider it as a real axis, which makes it possible to  assume $\pmb{\alpha}_0,\pmb{\alpha}_1$ as well as $k$ to be real numbers without loss of generality.

The optimal $k$ that maximizes $f(k):=|\pmb\alpha_0-\pmb\alpha_1|^2=(1+k)^2|\pmb\alpha_0|^2=\left((1+k)^2\E\right)/(1-q+k^2\cdot q)$  is equal to $k^*=(1-q)/q$.
By plugging this $k^*$ into~\eqref{eqn:a1a0k} and~\eqref{eqn:a1a0ke}, the optimal states are
\begin{equation}
\begin{split}\label{eqn:opta0a1p^*in}
\pmb\alpha_0^*&=\sqrt{\E \cdot q/(1-q)},\\
\pmb\alpha_1^*&=-\sqrt{\E \cdot(1-q)/{q}}.
\end{split}
\end{equation}
The maximum mutual information $C_{1,\sf{Binary}}(\E)$ with these optimal states can be written as
\begin{equation}\label{eqn:C1binref11}
C_{1,\sf{Binary}}(\E)=\max_{q} \left(H_B(q) -H_B(p^*(q))\right),
\end{equation}
for $0\leq q\leq1/2,$ where
\begin{equation}
\begin{split}\label{eqn:opta0a1p^*}
&p^*(q):=p(q,\pmb{\alpha_0}^*,\pmb{\alpha_1}^*)\\
&=\left(1-\sqrt{1-4q(1-q)\exp\left(-\frac{\E}{q(1-q)}\right)}\right)/2.
\end{split}
\end{equation}
Note that this $p^*(q)\leq q$ for every $\E\geq 0$. 

Now we want to find $q^*$ that maximizes the right hand side of \eqref{eqn:C1binref11}, which is defined as $I(q):=H_B(q) -H_B(p^*(q))$. 
The derivative of $I(q)$ is
\begin{equation}
\begin{split}\label{eqn:derIq}
&\frac{\partial I(q)}{\partial q}=\log\frac{1-q}{q}-\left(\log\frac{1-p^*(q)}{p^*(q)}\right)\left(\frac{1-2q}{1-2p^*(q)}\right)\times\\
&\qquad\qquad\qquad\qquad\left(1+\frac{\E}{q(1-q)}\right)\exp\left(-\frac{\E}{q(1-q)}\right).
\end{split}
\end{equation}
A closed form solution of $q^*$ that makes $\partial I(q)/\partial q|_{q=q^*}=0$ cannot be found, but instead we will show that for any $0\leq q\leq 1/2$,
\begin{equation}
I(q)\leq\E\log(1/\E)-\E\log\log(1/\E)+O(\E)
\end{equation}
as $\E\to 0$, and the equality can be met at $q=(\E/2)\log(1/\E)$, i.e.,
\begin{equation}
I(q)|_{q=\frac{\E}{2}\left(\log\frac{1}{\E}\right)}=\E\log(1/\E)-\E\log\log(1/\E)+O(\E).
\end{equation}
This will imply that 

\beq
C_{1,{\sf Binary}}(\E)=\E\log(1/\E)-\E\log\log(1/\E)+O(\E).
\eeq

Consider the following non-overlapping sub-intervals of $0\leq q\leq 1/2$:
\begin{enumerate}
\item $0\leq q < 0.9\E$,
\item $0.9\E \leq q <({\E}/{2})\sqrt{\log(1/\E)}$,
\item $({\E}/{2})\sqrt{\log(1/\E)}\leq q <({\E}/{2})({\log(1/\E)})^2$,
\item $({\E}/{2})({\log(1/\E)})^2\leq q <1/({\log(1/\E)})$,
\item ${1}/({\log(1/\E)})\leq q < {1}/{2}-{1}/{\sqrt{\log(1/\E)}}$,
\item ${1}/{2}-{1}/{\sqrt{\log(1/\E)}}\leq q\leq{1}/{2}$,
\end{enumerate}
for a sufficiently small $\E$.
We will show that the optimal $q^*$ that maximizes $I(q)$ and thus achieves $C_{1,\sf{Binary}}(\E)$ is in the sub-interval $3)$, and  $I(q)$ in the rest of the five sub-intervals are smaller than $C_{1,\sf{Binary}}(\E)$.

In the sub-interval 3), the probability $q\to 0$ as $\E\to 0$ and $p^*(q)$ in~\eqref{eqn:opta0a1p^*} can be approximated as
\beq
\begin{split}
&p^*(q)=q\exp\left(-\frac{\E}{q(1-q)}\right)+O(q^2)\\
&=q\left(1-\E/q+\E^2/(2q^2)\right)+O(\E^3/q^2).
\end{split}
\eeq
Using this and the approximation of the binary entropy $H_{\sf B}(x)=-x\log x+x+O(x^2)$ as $x\to 0$,
\begin{equation}
H_B(p^*(q))=-q\log q+\E\log q -(\E^2\log q)/(2q)+q+O(\E),
\end{equation}
and thus
\begin{equation}
\begin{split}
I(q)&=H_B(q)-H_B(p^*(q))\\
&=-\E\log q +(\E^2\log q)/(2q)+O(\E).
\end{split}
\end{equation}
By writing $q$ in this sub-interval as $q=\frac{\E}{2}\left(\log\frac{1}{\E}\right)^{\alpha}$ with a parameter $\alpha$ varying in $1/2\leq \alpha\leq 2$, 
\begin{equation}
\begin{split}\label{eqn:Iqlast}
I(q)&=\E\log\frac{1}{\E}-{\E}\left(\log\frac{1}{\E}\right)^{1-\alpha}-\alpha\E\log\log\frac{1}{\E}+O(\E).
\end{split}
\end{equation}
The derivative of $I(q)$ in $\alpha$ is
\begin{equation}
\partial I(q)/\partial \alpha=-\E\log\log\frac{1}{\E}\left(1-\left(\log\frac{1}{\E}\right)^{1-\alpha}\right).
\end{equation}
Since $\partial I(q)/\partial \alpha\leq 0$ for $1/2\leq \alpha \leq 1$, $\partial I(q)/\partial \alpha\geq 0$ for $1\leq \alpha \leq 2$ and  $\partial I(q)/\partial \alpha =0$ when $\alpha=1$, the optimal $q$ maximizing $I(q)$ is $q^*=\frac{\E}{2}\left(\log\frac{1}{\E}\right)$. At $\alpha=1$, $I(q)$ in (\ref{eqn:Iqlast}) becomes
\begin{equation}
\begin{split}
I(q)|_{q=\frac{\E}{2}\left(\log\frac{1}{\E}\right)}&=\E\log\frac{1}{\E}-\E\log\log\frac{1}{\E}+O(\E).
\end{split}
\end{equation}

We next show that in the rest of the five sub-intervals,
\begin{equation}\nonumber
I(q)\leq\E\log(1/\E)-\E\log\log(1/\E)+O(\E).
\end{equation}

In the first sub-interval of $0\leq q < 0.9\E$,
\begin{equation}
\begin{split}
&I(q)=H_B(q)-H_{\sf B}(p^*(q))\\
&\leq H_{\sf B}(q) \leq H_{\sf B}(q)|_{q=0.9\E}\\
&=0.9\E\log\frac{1}{\E}+O(\E)\\
&<  \E\log(1/\E)-\E\log\log(1/\E)+O(\E).
\end{split}
\end{equation}

To bound $I(q)$ for the rest of the four sub-intervals, we will use the mean value theorem, which shows that there exists a $r\in[p^*(q),q]$ satisfying
\begin{equation}\label{eqn:MVT}
I(q)=H_{\sf B}(q)-H_{\sf B}(p^*(q))=\left(\frac{\partial}{\partial p}H_{\sf B}(p)\right){\Bigg|}_{p=r}(q-p^*(q))
\end{equation}
where
\begin{equation}
\frac{\partial}{\partial p}H_{\sf B}(p)=\log\left(\frac{1}{p}-1\right).
\end{equation}
Since the derivative of entropy $H_{\sf B}(p)$ is a decreasing function in $0\leq p\leq1/2$ and $p^*(q)\leq q$ by the definition of $p^*(q)$ in (\ref{eqn:opta0a1p^*}),
\begin{equation}
\begin{split}
&\left(\frac{\partial}{\partial p}H_{\sf B}(p)\right){\Bigg|}_{p=r\in[p^*(q),q]}\\
&\leq \left(\frac{\partial}{\partial p}H_{\sf B}(p)\right){\Bigg|}_{p=p^*(q)}=\log\left(\frac{1}{p^*(q)}-1\right).
\end{split}
\end{equation}
We next find an upper bound on $\log\left({1}/{p^*(q)}-1\right)$ and an upper bound on $(q-p^*(q))$ in each sub-interval to show that $I(q)$ in~\eqref{eqn:MVT} is smaller than $C_{1,\sf Binary}(\E)$.

In the second sub-interval of $0.9\E \leq q <({\E}/{2})\sqrt{\log(1/\E)}$, $p^*(q)$ in~\eqref{eqn:opta0a1p^*} can be approximated as
\begin{equation}
\begin{split}\label{eqn:from81}
p^*(q)&=q\exp\left(-\frac{\E}{q(1-q)}\right)+O(q^2),
\end{split}
\end{equation}
and thus
\begin{equation}
q-p^*(q)=q\left(1-\exp\left(-\frac{\E}{q(1-q)}\right)\right)+O(q^2).
\end{equation}
By using $1-\exp(-x)\leq x-x^2/2+x^3/6$ for $x\geq0$, 
\begin{equation}
\begin{split}\label{eqn:q-pR2}
q-p^*(q)&\leq \frac{\E}{(1-q)}-\frac{\E^2}{2q(1-q)^2}+\frac{\E^3}{6q^2(1-q)^3}+O(q^2)\\
&= \E-\frac{\E^2}{2q}+\frac{\E^3}{6q^2}+O\left(\E q+q^2\right).
\end{split}
\end{equation}
Moreover, from~\eqref{eqn:from81}, 
\begin{equation}
{1}/{p^*(q)}\leq ({1}/{q})\left(1+O\left({\E}/{q}\right)\right).
\end{equation}
By using this, it can be shown that
\begin{equation}\label{eqn:derHpR2}
\begin{split}
&\left(\frac{\partial}{\partial p}H_{\sf B}(p)\right){\Bigg|}_{p=r\in[p^*(q),q]}\\
&\leq\log\left({1}/{p^*(q)}-1\right)\\
&\leq\log({1}/{q})+O\left( {\E}/{q}\right)=\log({1}/{q})+O(1)\\
&=\log({1}/{\E})+\log({\E}/{q})+O(1).
\end{split}
\end{equation}

 By using (\ref{eqn:q-pR2}),~\eqref{eqn:derHpR2} and~\eqref{eqn:MVT},
\begin{equation}\label{eqn:IqubR2}
\begin{split}
&I(q)\leq \E\log\frac{1}{\E}+\E\log\frac{\E}{q}-\frac{\E^2}{2q}\left(1-\frac{\E}{3q}\right)\log\frac{1}{\E}\\
&\quad\quad\;\;-\frac{\E^2}{2q}\left(1-\frac{\E}{3q}\right)\log\frac{\E}{q}+O\left(\E\right).
\end{split}
\end{equation}
In the interval $0.9\E \leq q <\frac{\E}{2}\sqrt{\log(1/\E)}$, the second term in the right hand side of (\ref{eqn:IqubR2}) can be bounded as
\begin{equation}
\E\log({\E}/{q})<\E.
\end{equation}
Moreover, since
$
0.6<\left(1-{\E}/({3q})\right)<1,
$
it can be shown that
\begin{equation}\label{eqn:Iqub1R2}
\begin{split}
&I(q)\\
&\leq \E\log\frac{1}{\E}-0.3\frac{\E^2}{q}\log\frac{1}{\E}-\frac{\E^2}{2q}\left(1-\frac{\E}{3q}\right)\log\frac{\E}{q}+O(\E)\\
&= \E\log\frac{1}{\E}-0.2\frac{\E^2}{q}\log\frac{1}{\E}\\
&\quad-0.1\frac{\E^2}{q}\left(\log\frac{1}{\E}+5\left(1-\frac{\E}{3q}\right)\log\frac{\E}{q} \right)+O(\E)
\end{split}
\end{equation}
Note that the term in the parenthesis of the right hand side is positive, i.e.,
\begin{equation}
\begin{split}
&\left(\log\frac{1}{\E}+5\left(1-\frac{\E}{3q}\right)\log\frac{\E}{q} \right)\\
&>\log\frac{1}{\E}+5\left(1-\frac{\E}{3q}\right)\left(-\frac{1}{2}\log\log\frac{1}{\E}\right)\\
&>\log\frac{1}{\E}-2.5\log\log\frac{1}{\E}>0
\end{split}
\end{equation}
as $\E\to 0$.
By using this fact, $I(q)$ in~\eqref{eqn:Iqub1R2} can be further bounded above as
\begin{equation}\label{eqn:Iqub2R2}
\begin{split}
I(q)
&\leq  \E\log\frac{1}{\E}-0.2\frac{\E^2}{q}\log\frac{1}{\E}+O(\E)\\
&\leq   \E\log\frac{1}{\E}-0.4{\E}\sqrt{\log\frac{1}{\E}}+O(\E),\\
&<\E\log\frac{1}{\E}-\E\log\log\frac{1}{\E}+O(\E).
\end{split}
\end{equation}

In the fourth sub-interval of $({\E}/{2})({\log(1/\E)})^2\leq q <{1}/({\log(1/\E)})$, by using
\begin{equation}
\begin{split}
&1-4q(1-q)\exp\left(-\frac{\E}{q(1-q)}\right)\\
&=1-4q(1-q)+4\E+O\left(\frac{\E^2}{q}\right)\\
&=(1-2q)^2\left(1+\frac{4\E}{(1-2q)^2}+O\left(\frac{\E^2}{q}\right)\right),
\end{split}
\end{equation}
it can be shown that $p^*(q)$ in~\eqref{eqn:opta0a1p^*} is
\begin{equation}
\begin{split}
p^*(q)&=\frac{1}{2}\left(1-(1-2q)\left(1+\frac{2\E}{(1-2q)^2}+O\left(\frac{\E^2}{q}\right)\right)\right)\\
&=q-\frac{\E}{1-2q}+O\left(\frac{\E^2}{q}\right)\\
&=q-\frac{\E}{1-2q}+O\left(\frac{\E}{(\log(1/\E))^2}\right).
\end{split}
\end{equation}
Then, $(q-p^*(q))$ can be bounded as
\begin{equation}
\begin{split}\label{eqn:q-p*R4}
q-p^*(q)&=\frac{\E}{1-2q}+O\left(\frac{\E}{(\log(1/\E))^2}\right)\\
&\leq \frac{\E}{1-\frac{2}{\log(1/\E)}}+O\left(\frac{\E}{(\log(1/\E))^2}\right)\\
&=\E+\frac{2\E}{\log(1/\E)}+O\left(\frac{\E}{(\log(1/\E))^2}\right).
\end{split}
\end{equation}
Moreover, in this region,
\begin{equation}\label{eqn:derHpR4}
\begin{split}
&\left(\frac{\partial}{\partial p}H_{\sf B}(p)\right){\Bigg|}_{p=r\in[p^*(q),q]}\\
&\leq\log\left(\frac{1}{p^*(q)}-1\right)=\log\left(\frac{1}{q\left(1-\frac{q-p^*(q)}{q}\right)}-1\right)\\
&\leq \log\frac{1}{q}+O\left(\frac{q-p^*(q)}{q}+q\right)\\
&\leq \log\frac{2}{\E\left(\log(1/\E)\right)^2}+o(1).
\end{split}
\end{equation}
From (\ref{eqn:q-p*R4}), (\ref{eqn:derHpR4}) and~\eqref{eqn:MVT},
\begin{equation}
\begin{split}
I(q)
&\leq  \left(\log\frac{2}{\E}-2\log\log\frac{1}{\E} \right)\left(\E+\frac{2\E}{\log(1/\E)}\right)+o(\E)\\
&\leq \E\log\frac{1}{\E}-2\E\log\log\frac{1}{\E}+O(\E),\\
&<\E\log\frac{1}{\E}-\E\log\log\frac{1}{\E}+O(\E).
\end{split}
\end{equation}

In the fifth sub-interval of ${1}/({\log(1/\E)})\leq q <{1}/{2}-{1}/{\sqrt{\log(1/\E)}}$, by using
\begin{equation}
\begin{split}
&1-4q(1-q)\exp\left(-\frac{\E}{q(1-q)}\right)\\
&=1-4q(1-q)+4\E+O\left(\frac{\E^2}{q}\right)\\
&=(1-2q)^2\left(1+\frac{4\E}{(1-2q)^2}+O\left(\frac{\E^2}{q(1-2q)^2}\right)\right),
\end{split}
\end{equation}
it can be shown that $p^*(q)$ in~\eqref{eqn:opta0a1p^*} is
\begin{equation}
\begin{split}
p^*(q)
&=q-\frac{\E}{1-2q}+O\left(\frac{\E^2}{(1-2q)^4}+\frac{\E^2}{q(1-2q)^2}\right).
\end{split}
\end{equation}
Moreover, in this sub-interval,
\begin{equation}
O\left(\frac{\E^2}{(1-2q)^3}+\frac{\E^2}{q(1-2q)^2}\right)=O\left(\E^2\left(\log(1/\E)\right)^{3/2}\right).
\end{equation}
Thus, the difference between $q$ and $p^*(q)$ in this region can be bounded as
\begin{equation}
\begin{split}\label{eqn:q-p*R5}
q-p^*(q)&=\frac{\E}{1-2q}+O\left(\E^2\left(\log(1/\E)\right)^{3/2}\right)\\
&\leq  \frac{1}{2}\E\sqrt{\log(1/\E)}+O\left(\E^2\left(\log(1/\E)\right)^{3/2}\right).
\end{split}
\end{equation}
Moreover,
\begin{equation}\label{eqn:derHpR5}
\begin{split}
&\left(\frac{\partial}{\partial p}H_{\sf B}(p)\right){\Bigg|}_{p=r\in[p^*(q),q]}\\
&\leq\log\left(\frac{1}{p^*(q)}-1\right)=\log\left(\frac{1}{q\left(1-\frac{q-p^*(q)}{q}\right)}-1\right)\\
&\leq \log\frac{1}{q}+O\left(\frac{q-p^*(q)}{q}+q\right)<\log\log(1/\E)+O(1).
\end{split}
\end{equation}
From (\ref{eqn:q-p*R5}), (\ref{eqn:derHpR5}) and~\eqref{eqn:MVT}, in the fifth sub-interval,
\begin{equation}
\begin{split}
I(q)
&\leq \frac{1}{2}\E\sqrt{\log(1/\E)}\left(\log\log(1/\E)\right)+O\left(\E\sqrt{\log(1/\E)}\right),\\
&<\E\log\frac{1}{\E}-\E\log\log\frac{1}{\E}+O(\E).
\end{split}
\end{equation}

Finally, we consider the sixth sub-interval,  ${1}/{2}-{1}/{\sqrt{\log(1/\E)}}\leq q\leq{1}/{2}$. When we denote $q:=1/2-\delta$ for $0\leq \delta\leq {1}/{\sqrt{\log(1/\E)}}$, by using
\begin{equation}
\begin{split}
&1-4q(1-q)\exp\left(-\frac{\E}{q(1-q)}\right)\\
&\leq 1-(1-4\delta^2)\left(1-\frac{4\E}{(1-4\delta^2)}\right)=4\delta^2+4\E,
\end{split}
\end{equation}
which is from $e^{-x}\geq 1-x$, it can be shown that  $p^*(q)$ in~\eqref{eqn:opta0a1p^*} is 
\begin{equation}
\begin{split}
p^*(q)
&\geq \frac{1}{2}\left(1-\sqrt{4\delta^2+4\E}\right)\\
&=q-\left(\frac{1}{2}-\delta\right)+\frac{1}{2}\left(1-\sqrt{4\delta^2+4\E}\right)\\
&=q+\left(\delta-\frac{1}{2}\sqrt{4\delta^2+4\E} \right).
\end{split}
\end{equation}
From this, we can write an upper bound on
\begin{equation}\label{eqn:q-p*R6}
q-p^*(q)\leq \frac{1}{2}\sqrt{4\delta^2+4\E} -\delta
\end{equation}
Moreover,
\begin{equation}\label{eqn:derHpR6}
\begin{split}
&\left(\frac{\partial}{\partial p}H_{\sf B}(p)\right){\Bigg|}_{p=r\in[p^*(q),q]}\\
&\leq\log\left(\frac{1}{p^*(q)}-1\right)\\
&\leq \log\left(\frac{2}{1-\sqrt{4\delta^2+4\E}}-1\right)\\
&=\log\left(1+2\sqrt{4\delta^2+4\E}+O(4\delta^2+4\E) \right)\\
&\leq 2\sqrt{4\delta^2+4\E}+O(4\delta^2+4\E) 
\end{split}
\end{equation}
where the last inequality is from $\log(1+x)\leq x$.

By combining (\ref{eqn:q-p*R6}), (\ref{eqn:derHpR6}) and~\eqref{eqn:MVT},
\begin{equation}
\begin{split}
I(q)&\leq \left(2\sqrt{4\delta^2+4\E}\right)\left(\frac{1}{2}\sqrt{4\delta^2+4\E} -\delta\right) \\
&\quad+O\left( \left(4\delta^2+4\E\right)   \left(\frac{1}{2}\sqrt{4\delta^2+4\E} -\delta\right)  \right)\\
&\leq  4\delta^2+4\E-2\delta\sqrt{4\delta^2+4\E}\\
&\quad+O\left( \left(4\delta^2+4\E\right)   \left(\frac{1}{2}\sqrt{4\delta^2+4\E} -\delta\right)  \right)\\
&=2\delta\left(2\delta-\sqrt{4\delta^2+4\E}\right)+4\E\\
&\quad+O\left( \left(4\delta^2+4\E\right)   \left(\frac{1}{2}\sqrt{4\delta^2+4\E} -\delta\right)  \right).
\end{split}
\end{equation}
Since $\left(2\delta-\sqrt{4\delta^2+4\E}\right)\leq 0$, 
\begin{equation}
2\delta\left(2\delta-\sqrt{4\delta^2+4\E}\right)+4\E\leq 4\E.
\end{equation}
By using this,
\begin{equation}
\begin{split}
I(q)&\leq 4\E+o(\E),\\
&<\E\log\frac{1}{\E}-\E\log\log\frac{1}{\E}+O(\E).
\end{split}
\end{equation}

In summary, we showed that in all the sub-intervals of $0\leq q\leq 1/2$, $I(q)\leq \E\log\frac{1}{\E}-\E\log\log\frac{1}{\E}+O(\E)$, and the equality is achieved at $q^*=\frac{\E}{2}\log\frac{1}{\E}$.

\section{Proof of Theorem \ref{cor1}}\label{app:cor1}
From Theorem \ref{thm:main1}, $C_N/N$ is bounded below as
\begin{equation}\label{eqn:main1r}
\frac{C_N}{N}\geq \max_R\left( \left(1-2e^{-N E(R)}\right)R-\frac{\log 2}{N}\right),
\end{equation}
where 
\begin{equation}
E(R)=\max_{0\leq s \leq 1}\left( \max_{P_X}\left(-\log\Tr(\rho^{1+s})\right)-sR\right),
\end{equation}
with $\rho=\sum_{x \in {\mathcal X}} P_X(x) \ket{\psi_x}\bra{\psi_x}$. We use this result to derive a lower bound on $C_{N,{\sf BPSK}}(\E)/N$ for the BPSK $\{\ket{\sqrt{\pmb\E}},\ket{-\sqrt{\pmb\E}}\}$ cq channel. Note that the inner product between the two states is equal to $\gamma=|\braket{{\sqrt{\pmb\E}}}{{-\sqrt{\pmb\E}}}|=e^{-2\E}$.

Let us first analyze the error exponent $E(R)$ of this cq channel. 
For the BPSK quantum states $\{\ket{\sqrt{\pmb\E}},\ket{-\sqrt{\pmb\E}}\}$ with input probabilities $\{1-q,q\}$, the two eigenvalues of the resulting density operator $\rho=(1-q)\ket{\sqrt{\pmb\E}}\bra{\sqrt{\pmb\E}}+q\ket{-\sqrt{\pmb\E}}\bra{-\sqrt{\pmb\E}}$ are 
\begin{equation}
\begin{split}\label{eqn:eignen_BPSK}
\sigma_1&=\left(1-\sqrt{1-4q(1-q)(1-e^{-4\E})}\right)/2,\\
\sigma_2&=\left(1+\sqrt{1-4q(1-q)(1-e^{-4\E})}\right)/2,
\end{split}
\end{equation}
as shown in (\ref{eqn:eigenvalues}).
It can be easily checked  that the optimal $q$ that maximizes
\begin{equation}
-\log\Tr(\rho^{1+s})=-\log\left(\sigma_1^{1+s}+\sigma_2^{1+s}\right)
\end{equation}
is equal to $1/2$.
When $\sigma_1$ and $\sigma_2$ at $q=1/2$ are denoted as $\sigma_1^*$ and $\sigma_2^*$, respectively,
\begin{equation}
\begin{split}\label{eqn:eignen_BPSK*}
\sigma_1^*&=(1-e^{-2\E})/2,\\
\sigma_2^*&=(1+e^{-2\E})/2.
\end{split}
\end{equation}
Then, the error exponent $E(R)$ for the BPSK inputs can be written in terms of $\sigma_1^*$ and $\sigma_2^*$ as
\begin{equation}
\begin{split}\label{eqn:opts01}
E(R)=&\max_{0\leq s \leq 1}\left( \max_{P_X}\left(-\log\Tr(\rho^{1+s})\right)-sR\right)\\
=&\max_{0\leq s \leq 1} \left(-\log\left((\sigma_1^*)^{1+s}+(\sigma_2^*)^{1+s}\right)-sR\right).
\end{split}
\end{equation}
To write $E(R)$ in terms of the mean photon number $\E$ and the rate $R$, we need to find the solution for the optimization in the right hand side of~\eqref{eqn:opts01} over ${0\leq s \leq 1} $. But, a closed form solution for $s$ cannot be found. Instead, by choosing $s^*$ from the approximation of the optimization as $\E\to0$,
we can find a lower bound on $E(R)$, denoted as $\widetilde{E}(R)$,
\begin{equation}\label{eqn:tildeER}
E(R)\geq \widetilde{E}(R):= \left(-\log\left((\sigma_1^*)^{1+s^*}+(\sigma_2^*)^{1+s^*}\right)-s^* R\right),
\end{equation}
where
\begin{equation}\label{eqn:s'three} 
\begin{split}
s^*:= \left\{ 
  \begin{array}{l l}  
    \frac{\log\log(1/\E)-\log(R-\E)}{\log(1/\E)}-1, & R_c \leq R \leq C, \\
    1, & R< R_c,\\
    0, & R> C,
  \end{array} \right.
\end{split}
\end{equation}
for $R_c:=\E+\E^2\log(1/\E)$ and $C:=\E\log(1/\E)+\E$. Note that in $R_c \leq R \leq C $, the defined $s^*$ is in $0\leq s^*\leq 1$.

Since $E(R)\geq\widetilde{E}(R)$ for every $R>0$, the lower bound on $C_N/N$ in (\ref{eqn:main1r}) can be further bounded below by using $\widetilde{E}(R)$ as follows,
\begin{equation}
\begin{split}\label{eqn:newCnlwbd}
\frac{C_{N,\sf{BPSK}}(\E)}{N}
&\geq\max_R \left( (1-2e^{-N \widetilde{E}(R)})R-\frac{\log 2}{N}\right).
\end{split}
\end{equation}
A closed form solution for the optimal $R$ that maximizes the lower bound in (\ref{eqn:newCnlwbd}) cannot be found. Instead, we choose
\begin{equation}\label{eqn:R**}
R^*=\E\log\frac{1}{\E}\left(1-\sqrt{\frac{\log\left(N\E\log(N\E)\right)}{N\E}}\right)+\E
\end{equation}
in the region of the blocklength $N\geq\E^{-1}\log(1/\E)$. We will show that for $\E\leq e^{-2}\approx 0.13$, the chosen rate $R^*$ is in $R_c\leq R^*\leq C$ when $N\geq\E^{-1}\log(1/\E)$. This implies that, at $R=R^*$, $s^*$ in~\eqref{eqn:s'three} belongs to the first case. 
To show this, we use the fact that when $N\E\geq 2$,
\begin{equation}\label{eqn:NEcond1}
0\leq \sqrt{\frac{\log\left(N\E\log(N\E)\right)}{N\E}}\leq 0.85,
\end{equation}
which can be validated by numerical computations using a computer. Under the assumption of $N\geq\E^{-1}\log(1/\E)$, if $\log(1/\E)\geq 2$, i.e., $\E\leq e^{-2}$, then $N\E\geq 2$.
Therefore, when $N\geq\E^{-1}\log(1/\E)$ and $\E\leq e^{-2}$, $R^*$ in (\ref{eqn:R**}) is in the range of
\begin{equation}
0.15\left(\E\log\frac{1}{\E}\right)+\E \leq R^*\leq \E\log\frac{1}{\E}+\E.
\end{equation}
Moreover, since $\E\leq e^{-2}< 0.15$,
\begin{equation}
\E+\E^2\log(1/\E) \leq 0.15\left(\E\log\frac{1}{\E}\right)+\E \leq R^*\leq \E\log\frac{1}{\E}+\E,
\end{equation}
and thus $R_c\leq R^*\leq C$.
 
In summary, for $\E\leq e^{-2}$ and $N\geq\E^{-1}\log(1/\E)$,
\begin{equation}\label{eqn:cor_ineq1}
\frac{C_{N,\sf{BPSK}}(\E)}{N}\geq  (1-2e^{-N \widetilde{E}(R^*)})R^*-\frac{\log 2}{N}.
\end{equation}
Furthermore, by numerical calculations, it can be shown that the lower bound in (\ref{eqn:cor_ineq1}) strictly increases with $N$ if $\E\leq 0.01$. 

\section{Proof of Corollary \ref{cornew1} }\label{app:cornew1}
The result in Corollary \ref{cornew1} can be derived by approximating the lower bound in Theorem \ref{cor1} under the assumption of $\E\to 0$. 
Let us first find the approximation of $\widetilde{E}(R)$ in (\ref{eqn:tildeER}).
For $0<s<1$, by using the Taylor expansions,
\begin{equation}
\begin{split}
({\sigma_1^*})^{1+s}&=((1-e^{-2\E})/2)^{1+s}=\E^{1+s}+O(\E^{2+s}),\\
({\sigma_2^*})^{1+s}&=
((1+e^{-2\E})/2)^{1+s}=1-(1+s)\E+O(\E^2),
\end{split}
\end{equation}
as $\E\to0$.
By using these approximations and the Taylor expansion of $\log(1+x)=x+O(x^2)$ as $x\to0$, 
\begin{equation}
-\log\left((\sigma_1^*)^{1+s}+(\sigma_2^*)^{1+s}\right)=(1+s)\E-\E^{1+s}+O(\E^2).
\end{equation}
Then, for $s=s^*$ in (\ref{eqn:s'three}), in the range of $R_c\leq R \leq C$,
\begin{equation}
\begin{split}
\widetilde{E}(R)&=(1+s^*)\E-\E^{1+s^*}-s^*R+O(\E^2)\\
&=\frac{(R-\E)}{\log(1/\E)}\left(\log(R-\E)+\log\frac{1}{\E}-\log\log\frac{1}{\E}-1\right)\\
&\quad+\E+O(\E^2).
\end{split}
\end{equation}
Now, at $R=R^*$ in \eqref{eqn:R**}, which was shown to be $R_c\leq R^* \leq C$ in Appendix \ref{app:cor1},
\begin{equation}
\begin{split}\label{eqn:127Etilde}
\widetilde{E}(R^*)&=\E\cdot\left(\sqrt{f(N,\E)}+\log\left(1-\sqrt{f(N,\E)}\right)\right.\\
&\left.\quad\qquad-\sqrt{f(N,\E)}\log\left(1-\sqrt{f(N,\E)}\right)\right)+O(\E^2)
\end{split}
\end{equation}
where
\begin{equation}
f(N,\E):=\frac{\log\left(N\E\log(N\E)\right)}{N\E}.
\end{equation}
In the range of $N\geq \E^{-1}\log(1/\E)$, i.e., $N\E \geq \log(1/\E)$, as $\E\to 0$ the resulting $N\E\to\infty$, and thus $f(N,\E)\to 0$. Therefore, in this regime of $N\geq \E^{-1}\log(1/\E)$, $\widetilde{E}(R^*)$ in~\eqref{eqn:127Etilde} can be further approximated as
\begin{equation}
\begin{split}\label{eqn:approx_f}
\widetilde{E}(R^*)&=(\E\cdot f(N,\E))/2+o(\E\cdot f(N,\E))+O(\E^2).
\end{split}
\end{equation}
If we further restrict the range of $N$ such that 
\begin{equation}\label{eqn:rangeofN}
\E^{-1}\log(1/\E)\leq N\leq\E^{-2}, \;{\text i.e., }\; \log(1/\E)\leq  N\E \leq  \E^{-1},\nonumber
\end{equation}
since $\E^2\leq 1/N$, $\widetilde{E}(R^*)$ becomes
\begin{equation}
\widetilde{E}(R^*)=\frac{\E}{2}\cdot\frac{\log\left(N\E\log(N\E)\right)}{N\E} +o\left(\frac{\log\left(N\E\log(N\E)\right)}{N} \right).
\end{equation}
Therefore, in the range of $\E^{-1}\log(1/\E)\leq N\leq\E^{-2}$,
\begin{equation}
\begin{split}
N\widetilde{E}(R^*)&=\log\sqrt{N\E\log(N\E)}+o(\log\sqrt{N\E\log(N\E)}),\\
e^{-N\widetilde{E}(R^*)}&=O\left(\frac{1}{\sqrt{N\E\log(N\E)}}\right).
\end{split}
\end{equation}
By using this result, the lower bound on ${C_{N,\sf{BPSK}}(\E)}/{N}$ in (\ref{eqn:cor_ineq1}) can be simplified as
\begin{equation}\label{eqn:CNfirstsimpa}
\begin{split}
\frac{C_{N,\sf{BPSK}}(\E)}{N}&\geq \E\log\frac{1}{\E}\left(1-\sqrt{\frac{\log\left(N\E\log(N\E)\right)}{N\E}}\right)\\
&\quad+\E+O\left(\frac{\E\log(1/\E)}{\sqrt{N\E\log(N\E)}}+\frac{\E}{\log(1/\E)}\right)
\end{split}
\end{equation}
in the range of $N$ such that $\E^{-1}\log(1/\E)\leq N\leq\E^{-2}$.
Moreover, in a narrower region of $N$ such that $\E^{-1}(\log(1/\E))^2\leq N \leq \E^{-2}$, the term $O\left(\frac{\E\log(1/\E)}{\sqrt{N\E\log(N\E)}}+\frac{\E}{\log(1/\E)}\right)$ can be simplified as $o(\E)$, and thus 
\begin{equation}\label{eqn:CNBPSKLBa}
\begin{split}
\frac{C_{N,\sf{BPSK}}(\E)}{N}&\geq \E\log\frac{1}{\E}\left(1-\sqrt{\frac{\log\left(N\E\log(N\E)\right)}{N\E}}\right)\\
&\quad+\E+o(\E).
\end{split}
\end{equation}

From (\ref{eqn:CNfirstsimpa}), it can be shown that at 
\begin{equation}\label{eqn:N11a}
N=2\E^{-1}\left(\log(1/\E)\right)^{2}\left(\log\log(1/\E)\right)^{-1},
\end{equation}
\begin{equation}\label{eqn:CNBPSKBa}
\frac{C_{N,\sf{BPSK}}(\E)}{N}\geq \E\log\frac{1}{\E}-\E\log\log\frac{1}{\E}+o\left(\E\log\log\frac{1}{\E}\right).
\end{equation}
Moreover, from (\ref{eqn:CNBPSKLBa}), it can be shown that at 
\begin{equation}\label{eqn:N12a}
N=\E^{-1}\left(\log(1/\E)\right)^{2}\left(\log\log(1/\E)\right)^2,
\end{equation} 
\begin{equation}\label{eqn:CNBPSKBPSKa}
\frac{C_{N,\sf{BPSK}}(\E)}{N}\geq \E\log\frac{1}{\E}+\E+o(\E).
\end{equation}



\bibliographystyle{IEEEtran}

\newpage
\begin{IEEEbiographynophoto}{Hye Won Chung}
(S'08--M'15) received the B.S. degree in Electrical Engineering with summa cum laude from the Korea Advanced Institute of Science and Technology (KAIST) in 2007, and the M.S. and Ph.D. degrees in Electrical Engineering and Computer Science from the Massachusetts Institute of Technology (MIT) in 2009 and 2014, respectively. Since 2014, she has been working as a Research Fellow in the Department of Electrical Engineering and Computer Science, University of Michigan. Her research interests include information theory, statistical inference, machine learning and quantum optical communications. 
Dr. Chung was awarded the Kwanjeong Educational Foundation Fellowship in 2007.
\end{IEEEbiographynophoto}

\begin{IEEEbiographynophoto}{Saikat Guha}
(M'09--SM'16) was born in Patna, India, in 1980. He received the Bachelor of Technology degree in Electrical Engineering from the Indian Institute of Technology (IIT) Kanpur, India in 2002, and the S.M. (Master of Science) and Ph.D. degrees in Electrical Engineering and Computer Science (EECS) from the Massachusetts Institute of Technology (MIT), Cambridge, MA in 2004 and 2008, respectively.

In 2008, he joined the Quantum Information Processing at BBN Technologies (now, Raytheon BBN Technologies), Cambridge, MA, as Scientist. He became a Senior Scientist in 2012, and a Lead Scientist in 2016. His current research interests include the application of quantum information and estimation theory to fundamental limits of optical communications and imaging, all-optical classical and quantum computing, quantum error correction, and network theory. Dr. Guha has led as principal investigator (PI) and served as co-PI, on several research programs on optics-based information processing and network theory, funded by the Defense Advanced Research Projects Agency (DARPA), Office of Naval Research (ONR), National Science Foundation (NSF), Sandia National Laboratory, and the Army Research Laboratory (ARL). He is a member of the Optical Society of America (OSA) and a Senior Member of IEEE.
 
He represented India at the 29th International Physics Olympiad at Reykjavik, in July 1998, where he received the European Physical Society award. He received the Raymie Stata Award for Outstanding Teaching in 2005 from the department of EECS of MIT. He was a co-recipient of a NASA Tech Brief Award in 2010, awarded by the NASA Inventions and Contributions Board, for his work on the phase-conjugate receiver for Gaussian-state quantum illumination. In 2011, he received the Raytheon Excellence in Engineering and Technology (EIET) Award, Raytheon's highest technical honor, for the exceptional scientific contributions of a DARPA Information in a Photon program team led by him. In 2013, he received the Anita Jones Entrepreneurial Award from BBN Technologies, in recognition of his work on quantifying the physical limits of optical communications and imaging, and developing novel techniques to approach these limits.
\end{IEEEbiographynophoto}

\begin{IEEEbiographynophoto}{Lizhong Zheng}
(S'00--M'02--F'16) received the B.S. and M.S. degrees, in 1994 and 1997 respectively, from the Department of Electronic Engineering, Tsinghua University, China, and the Ph.D. degree, in 2002, from the Department of Electrical Engineering and Computer Sciences, University of California, Berkeley. Since 2002, he has been working in the Department of Electrical Engineering and Computer Sciences, where he is currently a professor of electrical engineering. His research interests include information theory, statistical inference, wireless communications and networks. He is an IEEE fellow. He received the Eli Jury award from UC Berkeley in 2002, IEEE Information Theory Society Paper Award in 2003, and NSF CAREER award in 2004, and the AFOSR Young Investigator Award in 2007. 
\end{IEEEbiographynophoto}

\end{document}